\documentclass[12pt,twoside, a4paper]{article}
\def\pd{\partial}
\def\mc{\mathcal}

\usepackage[dvips]{graphicx}
\usepackage{amssymb}
\usepackage{amssymb,amsmath}
\usepackage{graphicx}
\usepackage{dsfont}
\usepackage{caption}
\usepackage{subcaption}
\usepackage{mathtools}
\usepackage{verbatim}
\usepackage{graphicx}
\usepackage{multirow}
\usepackage[outline]{contour}
\usepackage{xcolor,colortbl}
\input{epsf.sty}  
\begin{document}
\begin{center}
\Large{\textbf{Conformal line defects from matter-coupled 5D $N=4$ gauged supergravity}}
\end{center}
\vspace{1 cm}
\begin{center}
\large{\textbf{Parinya Karndumri}}
\end{center}
\begin{center}
String Theory and Supergravity Group, Department
of Physics, Faculty of Science, Chulalongkorn University, 254 Phayathai Road, Pathumwan, Bangkok 10330, Thailand \\
E-mail: parinya.ka@hotmail.com \vspace{1 cm}
\end{center}
\begin{abstract}
We study a number of holographic solutions describing conformal line defects within $N=2$ SCFTs from matter-coupled $N=4$ gauged supergravity in five dimensions. For $SO(2)_D\times SO(3)$ gauge group, the gauged supergravity admits two supersymmetric $AdS_5$ vacua with $N=4$ and $N=2$ unbroken supersymmetries. It turns out that only solutions that are asymptotic to the $N=4$ vacuum are possible leading to line defects within the dual $N=2$ SCFT. In gauged supergravity with $SO(2)\times ISO(3)$ gauge group, we find similar solutions with the $N=2$ SCFT appearing in the IR while the UV field theories correspond to six-dimensional $N=(0,2)$ field theories arising from M5-branes wrapped on a hyperbolic space $H^2$. The solution with $SO(2)\times SO(3)$ symmetry is also uplifted to eleven dimensions via a consistent truncation on $H^2\times S^4$. Finally, for gauged supergravity with $SO(2)\times SO(3)\times SO(3)$ gauge group that admits two $N=4$ supersymmetric $AdS_5$ vacua, we find a solution describing a line defect that interpolates between these two vacua as well as a solution with only one asymptotic $AdS_5$ geometry. 
\end{abstract}
\newpage
\section{Introduction}
The study of conformal defects of different dimensionalities within higher-dimensional conformal field theories has recently been attracted much attention in various areas of theoretical physics, see \cite{defect_review} for a review. For almost three decades, gauge/gravity duality, a generalization to non-conformal field theories of the AdS/CFT correspondence \cite{maldacena,Gubser_AdS_CFT,Witten_AdS_CFT}, has provided a very useful tool to study various aspects of strongly-coupled field theories. It is then natural to apply this holographic duality to study strongly coupled superconformal field theories (SCFTs) in the presence of conformal defects preserving lower-dimensional conformal subgroup of the conformal symmetry of the parent SCFTs, see \cite{defect1}-\cite{correlator_defect} for selected previous works along this line. 
\\
\indent In this paper, we are interested in holographic solutions describing one-dimensional or line defects within the dual $N=2$ SCFTs in four dimensions. The framework under consideration here is the half-maximal $N=4$ gauged supergravity in five dimensions coupled to a number of vector multiplets. The first example for this type of solutions has been found recently in pure $N=4$ gauged supergravity \cite{Romans_5DN4} by using an $AdS_2\times S^2$-sliced domain wall ansatz for the five-dimensional metric in \cite{line_defects_5DN4_Nicolo}. The $AdS_2$ and $S^2$ geometries are supported by non-vanishing two-form fields in the supergravity multiplet. In the present work, we aim to extend this result to matter-coupled $N=4$ gauged supergravity. We will consider three gauge groups, $SO(2)_D\times SO(3)$, $SO(2)\times ISO(3)$ and $SO(2)\times SO(3)\times SO(3)$. All of these gauged supergravities can be constructed by coupling the $N=4$ supergravity multiplet to three vector multiplets. 
\\
\indent The first two gauge groups have been shown to arise from consistent truncations of M-theory in \cite{Malek_AdS5_N4_embed}. In particular, the complete truncation ansatz of eleven-dimensional supergravity on $H^2\times S^4$ leading to $N=4$ gauged supergravity with $SO(2)\times ISO(3)$ gauge group has been given in \cite{ISO3_5D_N4_gauntlett}. Although the explicit truncation ansatz for $SO(2)_D\times SO(3)$ gauged supergravity is currently not known, this gauged supergravity admit two supersymmetric $AdS_5$ vacua with $N=4$ and $N=2$ supersymmetries. It is then interesting to look for line defect solutions interpolating between these $AdS_5$ vacua similar to the RG flow solutions studied in \cite{5D_N4_flow_Davide,5D_flowII}. On the other hand, the higher-dimensional origin of the $SO(2)\times SO(3)\times SO(3)$ gauge group is not known to date. However, it is still interesting to consider line defect solutions in this case since this gauged supergravity admits two $N=4$ supersymmetric $AdS_5$ vacua, see \cite{5D_N4_flow_Davide,5D_N4_flow} in which RG flow solutions between these vacua have been given. We also expect to find defect solutions interpolating between these vacua.  
\\
\indent Similar solutions describing line and surface defects within six- and five-dimensional SCFTs have been given in \cite{AdS2_F4_Dibitetto}-\cite{ISO3_defect} by using similar ansatze for the metric and other fields. As in the present case, the solutions have been found firstly in pure half-maximal gauged supergravity in \cite{AdS2_F4_Dibitetto,surface_defect_F4_Dibitetto,7D_sol_Dibitetto} and then extended to matter-coupled gauged supergravity in \cite{7D_N2_DW_3_form,F4_defect,ISO3_defect}. In the extension to matter-coupled theories, it turns out that coupling to vector multiplets puts severe constraints on the form of the solutions namely only charged domain walls are possible. In these solutions which take the form of domain walls with $AdS_2\times S^3$, $AdS_3\times S^2$ or $AdS_3\times S^3$ slices, the warp factors of the $AdS_{2,3}$ and the $S^{2,3}$ are the same, and the phase functions of the Killing spinors are trivial, being a constant. This is primarily due to the fact that the two- or three-form fields within the supergravity multiplet, which are relevant to support the $AdS_{2,3}\times S^{3,2}$ or $AdS_3\times S^3$ geometries on the world-volume, do not appear in the supersymmetry transformations of gauginos, see \cite{7D_N2_DW_3_form,F4_defect,ISO3_defect} for more detail. This results in solutions that are essentially of the same form as holographic RG flows with non-vanishing two- or three-form fields since the BPS equations for scalars from vector multiplets are the same. 
\\
\indent In the present case, we expect something different to occur since the supersymmetry transformations of the gauginos in five-dimensional gauged supergravity coupled to matter multiplets do indeed contain a contribution from the two-form fields \cite{N4_gauged_SUGRA,5D_N4_Dallagata}. However, as we will see, even this does not lead to line defect solutions different from charged domain walls. In this case, the manner in which the contributions of the two-form fields enter the BPS equations is rather restrictive such that these contributions must be zero whenever the scalar fields from the vector multiplets are non-vanishing. Although the solutions only take the form of charged domain walls, we do find interesting solutions some of which can be uplifted to eleven dimensions. On the other hand, there exists a line defect solution that interpolates between two $N=4$ $AdS_5$ vacua in $SO(2)\times SO(3)\times SO(3)$ gauge group which are related by a holographic RG flow.  
\\
\indent The paper is organized as follows. In section \ref{N4_SUGRA},
we give a brief review of five-dimensional $N=4$ gauged supergravity coupled to vector multiplets. In the following three sections \ref{SO2D_SO3}, \ref{SO2_ISO3} and \ref{SO2_SO3_SO3}, a number of supersymmetric line defects are given. In particular, with the known complete truncation ansatz of eleven-dimensional supergravity on $H^2\times S^4$, we perform an uplift of the $SO(2)\times SO(3)$ symmetric solution obtained from $SO(2)\times ISO(3)$ gauge group in section \ref{SO2_ISO3}. We end the paper by giving some conclusions and comments in section \ref{conclusion}. In the appendix, we present the detailed analysis of the BPS equations used in the main text. This might be useful to other related studies as well.

\section{Five-dimensional $N=4$ gauged supergravity}\label{N4_SUGRA}
In this section, we review the general structure of $N=4$ gauged supergravity in the embedding tensor formalism. We will mainly present the bosonic Lagrangian and supersymmetry transformations of fermionic fields and refer to \cite{N4_gauged_SUGRA,5D_N4_Dallagata} for more details.
\\
\indent The $N=4$ supergravity multiplet consists of the graviton
$e^{\hat{\mu}}_\mu$, four gravitinos $\psi_{\mu i}$, six vectors $(A^0_\mu,A_\mu^m)$, four spin-$\frac{1}{2}$ fields $\chi_i$ and one real scalar $\Sigma$, the dilaton. Space-time and tangent space indices are denoted respectively by $\mu,\nu,\ldots =0,1,2,3,4$ and
$\hat{\mu},\hat{\nu},\ldots=0,1,2,3,4$. The fundamental representation of $SO(5)_R\sim USp(4)_R$
R-symmetry is labeled by $m,n=1,\ldots, 5$ for $SO(5)_R$ and $i,j=1,2,3,4$ for $USp(4)_R$. The supergravity multiplet can couple to an arbitrary number $n$ of vector multiplets $(A^a_\mu,\lambda^{a}_i,\phi^{ma})$ with $a=1,\ldots, n$. Each vector multiplet, which is the only matter multiplet in $N=4$ supersymmetry, contains a vector field $A_\mu$, four gauginos $\lambda_i$ and five scalars $\phi^m$. All fermionic fields $\Psi_i=(\psi_{\mu i},\chi_i,\lambda^{a}_i)$ in the fundamental representation of $USp(4)_R$ are symplectic Majorana spinors subject to the condition
\begin{equation}
\Psi_i=\Omega_{ij}C(\bar{\Psi}^j)^T 
\end{equation}
with $C$ and $\Omega_{ij}$ being the charge conjugation and $USp(4)$ symplectic matrices, respectively.
\\
\indent In the matter-coupled $N=4$ supergravity, there are $6+n$ vector fields denoted collectively by $A^{\mc{M}}_\mu=(A^0_\mu,A^m_\mu,A^a_\mu)$ and $5n+1$ scalars in the $SO(1,1)\times SO(5,n)/SO(5)\times SO(n)$ coset manifold. We have introduced a collective index $\mc{M}=(0,M)$ as in \cite{N4_gauged_SUGRA}. The $5n$ scalars from the vector multiplets can be described by a coset representative $\mc{V}_M^{\phantom{M}A}$ transforming under the global $SO(5,n)$ and the local $SO(5)\times SO(n)$ respectively by left and right multiplications with the global $SO(5,n)$ indices $M,N,\ldots=1,2,\ldots , 5+n$ and the local $SO(5)\times SO(n)$ indices $A,B,\ldots$ split as $A=(m,a)$. Accordingly, the coset representative can be written as
\begin{equation}
\mc{V}_M^{\phantom{M}A}=(\mc{V}_M^{\phantom{M}m},\mc{V}_M^{\phantom{M}a}).
\end{equation}
It is also useful to define a symmetric and $SO(5)\times SO(n)$ invariant scalar matrix 
\begin{equation}
M_{MN}=\mc{V}_M^{\phantom{M}m}\mc{V}_N^{\phantom{M}m}+\mc{V}_M^{\phantom{M}a}\mc{V}_N^{\phantom{M}a}\, .
\end{equation}
\indent For $N=4$ gauged supergravity in five dimensions, the embedding of a gauge group in the global symmetry group $SO(1,1)\times SO(5,n)$ is characterized by the following components of the embedding tensor $\xi^{M}$, $\xi^{MN}=\xi^{[MN]}$ and $f_{MNP}=f_{[MNP]}$. In this paper, we will consider only gaugings with $\xi^M=0$ that admit supersymmetric $AdS_5$ vacua as shown in \cite{AdS5_N4_Jan}. In this case, the gauge group is embedded entirely in $SO(5,n)$ with the corresponding gauge generators in the fundamental representation of $SO(5,n)$ given by
\begin{equation}
{(X_M)_N}^P=-{f_M}^{QR}{(t_{QR})_N}^P={f_{MN}}^P\quad \textrm{and}\quad {(X_0)_N}^P=-\xi^{QR}{(t_{QR})_N}^P={\xi_N}^P\, .
\end{equation}
In these equations, $SO(5,n)$ generators are chosen to be ${(t_{MN})_P}^Q=\delta^Q_{[M}\eta_{N]P}$ with $\eta_{MN}=\textrm{diag}(-1,-1,-1,-1,-1,1,1,\ldots,1)$ being the $SO(5,n)$ invariant tensor. We also note the gauge covariant derivative
\begin{equation}
D_\mu=\nabla_\mu+A_\mu^{M}X_M+A^0_\mu X_0=\nabla_\mu+A^{\mc{M}}X_{\mc{M}}
\end{equation}
with $\nabla_\mu$ being a space-time covariant derivative including $SO(5)\times SO(n)$ composite connection.  
\\
\indent The bosonic Lagrangian of a general gauged $N=4$ supergravity can be written as
\begin{eqnarray}
e^{-1}\mc{L}&=&\frac{1}{2}R-\frac{3}{2}\Sigma^{-2}D_\mu \Sigma D^\mu \Sigma +\frac{1}{16} D_\mu M_{MN}D^\mu
M^{MN}-V\nonumber \\
& &-\frac{1}{4}\Sigma^2M_{MN}\mc{H}^M_{\mu\nu}\mc{H}^{N\mu\nu}-\frac{1}{4}\Sigma^{-4}\mc{H}^0_{\mu\nu}\mc{H}^{0\mu\nu}+e^{-1}\mc{L}_{\textrm{top}}
\end{eqnarray}
where $e$ is the vielbein determinant. $\mc{L}_{\textrm{top}}$ is a topological term describing the kinetic terms for two-form fields and the coupling between two-form and gauge fields. Since the explicit form of $\mc{L}_{\textrm{top}}$ is rather complicated, we will not give it here but refer to \cite{N4_gauged_SUGRA}. 
\\
\indent The covariant gauge field strength tensors read
\begin{equation}
\mc{H}^{\mc{M}}_{\mu\nu}=2\pd_{[\mu}A^{\mc{M}}_{\nu]}+{X_{\mc{N}\mc{P}}}^{\mc{M}}A^{\mc{N}}_\mu A^{\mc{P}}_\nu+Z^{\mc{M}\mc{N}}B_{\mu\nu\mc{N}}\label{covariant_field_strength}
\end{equation}
with 
\begin{equation}
Z^{MN}=\frac{1}{2}\xi^{MN}\qquad \textrm{and} \qquad Z^{0M}=-Z^{M0}=\frac{1}{2}\xi^M=0\, .
\end{equation}
\indent The two-form fields satisfy first-order field equations of the form 
\begin{equation}
Z^{\mc{M}\mc{N}}\left[\frac{1}{6\sqrt{2}}\epsilon_{\mu\nu\rho\lambda\sigma}\mc{H}^{(3)\rho\lambda\sigma}_{\mc{N}}-\mc{M}_{\mc{N}\mc{P}}
\mc{H}^{\mc{P}}_{\mu\nu}\right]=0\label{2-form_field_eq}
\end{equation}
with $\mc{M}_{00}=\Sigma^{-4}$, $\mc{M}_{0M}=0$ and $\mc{M}_{MN}=\Sigma^2M_{MN}$. This gives a duality relation between vectors and two-form fields with the three-form field strength $\mc{H}^{(3)}_{\mc{M}}$ defined by
\begin{equation}
Z^{\mc{M}\mc{N}}\mc{H}^{(3)}_{\mu\nu\rho\mc{N}}=Z^{\mc{M}\mc{N}}\left[3D_{[\mu}B_{\nu\rho]\mc{N}}
+6d_{\mc{NPQ}}A^{\mc{P}}_{[\mu}\left(\pd_\nu A^{\mc{Q}}_{\rho]}+\frac{1}{3}{X_{\mc{RS}}}^{\mc{Q}}A^{\mc{R}}_\nu A^{\mc{S}}_{\rho]}\right)\right]\label{H3_def}
\end{equation}
for $d_{0MN}=d_{MN0}=d_{M0N}=\eta_{MN}$ and 
\begin{equation}
{X_{MN}}^P={f_{MN}}^P,\qquad {X_{M0}}^0=0,\qquad {X_{0M}}^N={\xi_M}^N\, . 
\end{equation}
\indent Finally, the scalar potential is given by
\begin{eqnarray}
V&=&-\frac{1}{4}\left[f_{MNP}f_{QRS}\Sigma^{-2}\left(\frac{1}{12}M^{MQ}M^{NR}M^{PS}-\frac{1}{4}M^{MQ}\eta^{NR}\eta^{PS}\right.\right.\nonumber \\
& &\left.+\frac{1}{6}\eta^{MQ}\eta^{NR}\eta^{PS}\right) +\frac{1}{4}\xi_{MN}\xi_{PQ}\Sigma^4(M^{MP}M^{NQ}-\eta^{MP}\eta^{NQ})\nonumber \\
& &\left.
+\frac{\sqrt{2}}{3}f_{MNP}\xi_{QR}\Sigma M^{MNPQR}\right]
\end{eqnarray}
where $M^{MN}$ is the inverse of $M_{MN}$, and $M^{MNPQR}$ is obtained from
\begin{equation}
M_{MNPQR}=\epsilon_{mnpqr}\mc{V}_{M}^{\phantom{M}m}\mc{V}_{N}^{\phantom{M}n}
\mc{V}_{P}^{\phantom{M}p}\mc{V}_{Q}^{\phantom{M}q}\mc{V}_{R}^{\phantom{M}r}
\end{equation}
by raising the indices with $\eta^{MN}$. 
\\
\indent In order to find supersymmetric solutions, we also need supersymmetry transformations of fermionic fields. In the present analysis, it turns out that a slightly different form of supersymmetry transformations is more suitable. In particular, in contrast to \cite{N4_gauged_SUGRA}, we will explicitly show the $SO(5)_R$ gamma matrices ${(\Gamma_m)_i}^j$. In addition, we will also set all the vector fields to zero for simplicity since all the solutions considered here do not involve non-vanishing vector fields. With all these, the fermionic supersymmetry transformations are given by
\begin{eqnarray}
\delta\psi_{\mu i} &=&D_\mu \epsilon_i+\frac{i}{24}\Sigma {\mc{V}_M}^m{(\Gamma_m)_i}^j\xi^{MN}B_{N\nu\rho}({\gamma_\mu}^{\nu\rho}-4\delta^\nu_\mu\gamma^\rho)\epsilon_j\nonumber \\
& &+\frac{i}{\sqrt{6}}\gamma_\mu\left[\frac{1}{4\sqrt{3}}\Sigma^2\hat{\xi}^{mn}{(\Gamma_{mn})_i}^j-\frac{1}{6\sqrt{6}}\Sigma^{-1}\hat{f}_{mnp}{(\Gamma^{mnp})_i}^j\right]\epsilon_j,
\\
\delta \chi_i &=&-\frac{\sqrt{3}}{2}i\Sigma^{-1} D_\mu
\Sigma\gamma^\mu \epsilon_i+\frac{1}{8\sqrt{3}}\Sigma\xi^{MN}B_{N\mu\nu}\gamma^{\mu\nu}{\mc{V}_M}^m{(\Gamma_m)_i}^j\epsilon_j\nonumber \\
& &-\frac{1}{2\sqrt{6}}\Sigma^2\hat{\xi}^{mn}{(\Gamma_{mn})_i}^j\epsilon_j-\frac{1}{12\sqrt{3}}\Sigma^{-1}\hat{f}_{mnp}{(\Gamma^{mnp})_i}^j\epsilon_j,\\
\delta \lambda^a_i&=&P^a_{\mu m}{(\Gamma_m)_i}^j\epsilon_j-\frac{1}{4}\Sigma^{-1}\tilde{f}^{amn}{(\Gamma_{mn})_i}^j\epsilon_j-\frac{1}{2\sqrt{2}}\Sigma^2\tilde{\xi}^{am}{(\Gamma_m)_i}^j\epsilon_j\nonumber \\
& &-\frac{1}{8}\Sigma{\mc{V}_M}^a\xi^{MN}B_{N\mu\nu}\gamma^{\mu\nu}\epsilon_i
\end{eqnarray}
in which we have defined the ``dressed'' components of the embedding tensor by 
\begin{eqnarray}
& &\hat{\xi}^{mn}={\mc{V}_M}^m{\mc{V}_N}^n\xi^{MN},\qquad \hat{f}_{mnp}={\mc{V}_m}^M{\mc{V}_n}^N{\mc{V}_p}^Pf_{MNP},\nonumber \\
& &\tilde{\xi}^{am}={\mc{V}_M}^a{\mc{V}_N}^m\xi^{MN},\qquad \tilde{f}^{amn}={\mc{V}_M}^a{\mc{V}_N}^m{\mc{V}_P}^nf^{MNP}\, .
\end{eqnarray}
The vielbein on the coset manifold is defined by
\begin{equation}
P^a_{\mu m}=\frac{i}{2}{\mc{V}_M}^a\pd_\mu {\mc{V}_m}^M\, .
\end{equation}
\indent With vanishing vector fields, the covariant derivative on $\epsilon_i$ is simply given by
\begin{equation}
D_\mu \epsilon_i=\pd_\mu \epsilon_i+\frac{1}{4}\omega_\mu^{ab}\gamma_{ab}\epsilon_i+{\mc{V}_{ik}}^M\pd_\mu {\mc{V}_M}^{kj}\epsilon_j\, .
\end{equation}

\section{Line defects from $SO(2)_D\times SO(3)$ gauged supergravity}\label{SO2D_SO3}
We first consider holographic solutions describing line defects within $N=2$ four-dimensional superconformal field theories from $N=4$ gauged supergravity with $SO(2)_D\times SO(3)$ gauge group. We will consider $N=4$ gauged supergravity coupled to two vector multiplets with $SO(5,2)$ global symmetry. The gauge group is embedded in the compact subgroup $SO(5)_R\times SO(2)\subset SO(5,2)$ via $SO(2)_R\times SO(3)_R\times SO(2)\subset SO(5)_R\times SO(2)$ with $SO(2)_D$ being a diagonal subgroup of $SO(2)_R\times SO(2)$, and the $SO(3)$ factor is identified with $SO(3)_R\subset SO(5)_R$. This gauge group has been first considered in \cite{5D_N4_flow_Davide} in which it has been shown that the resulting gauged supergravity admits two supersymmetric $AdS_5$ vacua. One $AdS_5$ vacuum with $N=4$ supersymmetry preserves the full $SO(2)_D\times SO(3)$ symmetry while the other one with only $N=2$ supersymmetry is invariant under a smaller symmetry $SO(2)_{\textrm{diag}}\subset SO(2)_D\times SO(2)$ with $SO(2)\subset SO(3)$. Various holographic solutions from this gauged supergravity without two-form fields have been studied in a number of previous works \cite{5D_N4_flow_Davide,5D_flowII,5D_N4_Janus,5D_N4_black_stringII}. In this paper, we will look for supersymmetric $AdS_2\times S^2$-sliced domain wall solutions with non-vanishing two-form fields. These solutions are dual to line or one-dimensional conformal defects within the $N=2$ or $N=1$ SCFTs dual to the aforementioned $AdS_5$ vacua. In addition, this gauged supergravity has been shown to arise from a consistent truncation of eleven-dimensional supergravity on $M_3\times S^2\times S^1$ \cite{Malek_AdS5_N4_embed}, so in principle, it is possible to uplift the solutions from this gauged supergravity to M-theory.
\\
\indent In the convention of \cite{5D_N4_Janus}, the embedding tensor is given by
\begin{eqnarray}
\xi^{MN}&=&-g_1(\delta^M_1\delta^N_2-\delta^M_2\delta^N_1)+g_2(\delta^M_{6}\delta^N_{7}-\delta^M_{7}\delta^N_{6}),\\ 
f_{\tilde{m}+2,\tilde{n}+2,\tilde{p}+2}&=&-h_1\epsilon_{\tilde{m}\tilde{n}\tilde{p}},\qquad \tilde{m},\tilde{n},\tilde{p}=1,2,3
\end{eqnarray} 
with $g_1$, $g_2$ and $h_1$ being gauge coupling constants. To construct a coset representative of the $SO(5,n)/SO(5)\times SO(n)$ coset, we write the $SO(5,n)$ non-compact generators as
\begin{equation}
Y_{ma}=t_{m,a+5},\qquad m=1,2,\ldots, 5,\qquad a=1,2,\ldots, n
\end{equation}
with $n=2$ for the present case. We are interested in solutions preserving $SO(2)_{\textrm{diag}}$ symmetry given by the diagonal subgroup of $SO(2)_D$ and $SO(2)\subset SO(3)$. The corresponding coset representative can be written as, see more detail in \cite{5D_N4_Janus}, 
\begin{equation}
\mc{V}=e^{\phi_1\hat{Y}_1}e^{\phi_2\hat{Y}_2}e^{\phi_3\hat{Y}_3}e^{\phi_4\hat{Y}_4}\, .\label{coset_rep}
\end{equation}
with
\begin{eqnarray}
& &\hat{Y}_1=Y_{11}+Y_{22},\qquad \hat{Y}_2=Y_{12}-Y_{21},\nonumber \\
& & \hat{Y}_3=Y_{41}+Y_{52},\qquad \hat{Y}_4=Y_{42}-Y_{51}\, . 
\end{eqnarray}
As pointed out in \cite{5D_N4_Janus}, in order to consitently truncate out all the vector fields, we need to set $\phi_2=\phi_4=0$. 

\subsection{The analysis of BPS equations}
The metric ansatz is given by an $AdS_2\times S^2$-sliced domain wall of the form
\begin{equation}
ds^2=e^{2f(r)}ds^2_{AdS_2}+dr^2+e^{2h(r)}ds^2_{S^2}\, .
\end{equation}
We will split the five-dimensional coordinates as $x^\mu=(x^\alpha,r,x^s)$ for $\alpha=0,1$ and $s=3,4$. In order to preserve the $AdS_2\times S^2$ isometry, all the warp functions in the metric $(f,h)$ and scalar fields can depend only on the radial coordinate $r$. As in \cite{line_defects_5DN4_Nicolo}, the geometry is supported by non-vanishing two-form fields which, in the present case, are given by
\begin{equation}
B_M=b_M\textrm{vol}_{AdS_2}+\tilde{b}_M\textrm{vol}_{S^2}
\end{equation}
for $M=1,2,6,7$. All these two-forms are charged under the $SO(2)_D$ factor of the gauge group. We also note that in addition to $B_1$ and $B_2$ appearing in pure $N=4$ gauged supergravity, there are two extra two-form fields $B_6$ and $B_7$ arising from the dualization of the remaining two vector fields $A^6$ and $A^7$ that do not participate in the gauging. 
\\
\indent The corresponding three-form field strength tensors are given by
\begin{equation}
\mc{H}^{(3)}_M=dB_M=b_M'dr\wedge \textrm{vol}_{AdS_2}+\tilde{b}'_Mdr\wedge \textrm{vol}_{S^2}\, .
\end{equation}
We will use $'$ to denote $r$-derivatives throughout the paper. Vanishing of the gauge fields also implies the following constraint on the two-form fields of the form
\begin{equation}
d_{MNP}\mc{H}^M\wedge \mc{H}^N=d_{MNP}\xi^{MQ}\xi^{NR}B_Q\wedge B_R=0\, .
\end{equation}
For independent $g_1$ and $g_2$, we find that the above constraint gives
\begin{equation}
b_1\tilde{b}_1+b_2\tilde{b}_2=0\qquad \textrm{and}\qquad b_6\tilde{b_6}+b_7\tilde{b}_7=0\, .
\end{equation} 
These conditions can be solved by choosing
\begin{eqnarray}
b_2=kb_1,\qquad \tilde{b}_1=-k\tilde{b}_2\qquad \textrm{and}\qquad b_7=\tilde{k}b_6,\qquad \tilde{b}_6=-\tilde{k}\tilde{b}_7
\end{eqnarray} 
for constants $k$ and $\tilde{k}$. Follow \cite{line_defects_5DN4_Nicolo}, we can choose an appropriate $SO(2)_D$ transformation to set $k=\tilde{k}=0$ leading to
\begin{equation}
b_2=\tilde{b}_1=b_7=\tilde{b}_6=0\, .
\end{equation}
\indent Before considering the two-form field equations, we first note non-vanishing components of the scalar matrix $M_{MN}$ as follows
\begin{eqnarray}
& &M_{11}=M_{22}=\cosh^2\phi_1+\cosh2\phi_3\sinh^2\phi_1,\qquad M_{33}=M_{66}=1,\nonumber \\
& & M_{66}=M_{77}=\cosh^2\phi_1\cosh2\phi_3+\sinh^2\phi_1,\qquad M_{44}=M_{55}=\cosh2\phi_3,\nonumber \\
& &M_{17}=\cosh^2\phi_3\sinh2\phi_1,\qquad M_{41}=M_{52}=\sinh\phi_1\sinh2\phi_3,\nonumber \\
& & M_{47}=\cosh\phi_1\sinh2\phi_3\, .
\end{eqnarray}
With all these, the two-form field equations give rise to
\begin{eqnarray}
b'_1e^{2h-2f}-\frac{1}{\sqrt{2}}g_1\tilde{b}_2\Sigma^2(\cosh^2\phi_1+\sinh^2\phi_1)-\frac{1}{\sqrt{2}}g_2\tilde{b}_7\Sigma^2\sinh2\phi_1&=&0,\nonumber \\
\tilde{b}'_2e^{2f-2h}-\frac{1}{\sqrt{2}}g_1b_1\Sigma^2(\cosh^2\phi_1+\sinh^2\phi_1)-\frac{1}{\sqrt{2}}g_2b_6\Sigma^2\sinh2\phi_1&=&0,\nonumber \\
b'_6e^{2h-2f}-\frac{1}{\sqrt{2}}g_2\tilde{b}_7\Sigma^2(\cosh^2\phi_1+\sinh^2\phi_1)-\frac{1}{\sqrt{2}}g_1\tilde{b}_2\Sigma^2\sinh2\phi_1&=&0,\nonumber \\
\tilde{b}'_7e^{2f-2h}-\frac{1}{\sqrt{2}}g_2b_6\Sigma^2(\cosh^2\phi_1+\sinh^2\phi_1)-\frac{1}{\sqrt{2}}g_1b_1\Sigma^2\sinh2\phi_1&=&0\, .\label{2-form_eq}
\end{eqnarray}
\indent We now use the same ansatz for the Killing spinors as in \cite{line_defects_5DN4_Nicolo} of the form
\begin{equation}
\epsilon_i=Y\left[\cos\theta {\delta_i}^j+\sin\theta \gamma_{01}{(\Gamma_1)_i}^j\right]\tilde{\epsilon}_j\label{Killing_spin}
\end{equation}
in which $Y$ and $\theta$ are functions of $r$. We will follow \cite{line_defects_5DN4_Nicolo} and take the space-time gamma matrices to be
\begin{equation} 
 \gamma_{\hat{\alpha}}=\beta_\alpha,\qquad \gamma_{\hat{r}}=\beta_*\otimes \rho_*,\qquad \gamma_{\hat{s}}=\mathbb{I}_2\otimes \rho_{s-2}
\end{equation}
with $\beta_\alpha$ and $\rho_{s-2}$ are flat gamma matrices on $AdS_2$ and $S^2$, respectively. The explicit form for these matrices and the corresponding chirality matrices are given in terms of Pauli's matrices by
\begin{equation}
\beta_0=i\sigma^1,\qquad \beta_2=\sigma^2,\qquad \rho_1=\sigma^1,\qquad \rho_2=\sigma^2
\end{equation}
and 
\begin{equation}
\beta_*=-\beta_0\beta_1=\sigma^3,\qquad \rho_*=-i\rho_1\rho_2=\sigma^3\, .
\end{equation}
The spinor $\tilde{\epsilon}_j$ is written in terms of the Killing spinors on $AdS_2$ and $S^2$ as 
\begin{equation}
\tilde{\epsilon}_j=e^{i\varphi(r)\gamma_{\hat{r}}}\eta_j\otimes \chi_j
\end{equation}
with
\begin{equation}
\hat{\nabla}_\alpha\eta_i=\frac{\kappa_i}{2}\beta_\alpha\eta_i\qquad \textrm{and}\qquad \widetilde{\nabla}_s\chi_i=\frac{\tilde{\kappa}_i}{2}\rho_*\rho_s\chi_i\, .\label{AdS2_S2_Killing}
\end{equation}
In these equations, $\kappa_i=\pm 1$ and $\tilde{\kappa}_i=\pm 1$ are constants.
\\
\indent Finally, the explicit form of $SO(5)$ gamma matrices is chosen to be
\begin{eqnarray}
& &\Gamma_1=-\sigma^3\otimes \sigma^3,\qquad \Gamma_2=-\sigma^2\otimes \mathbb{I}_2,\qquad \Gamma_3=-\sigma^3\otimes \sigma^1,\nonumber \\
& &\Gamma_4=-\sigma^1\otimes \mathbb{I}_2,\qquad \Gamma_5=-\sigma^3\otimes \sigma^2
\end{eqnarray}     
such that $\Gamma_1\Gamma_2\Gamma_3\Gamma_4\Gamma_5=\mathbb{I}_4$. Before giving the BPS equations, we note all non-vanishing dressed components of the embedding tensor
\begin{eqnarray}
& &\tilde{\xi}^{12}=-\tilde{\xi}^{21}=-\frac{1}{2}(g_1-g_2)\cosh\phi_3\sinh2\phi_1,\nonumber \\
& &\tilde{\xi}^{15}=-\tilde{\xi}^{24}=\frac{1}{4}[g_1+g_2+(g_2-g_1)\cosh2\phi_1]\sinh2\phi_3,
\end{eqnarray}
\begin{eqnarray}
& &\hat{\xi}^{12}=-\frac{1}{2}[g_1+g_2+(g_1-g_2)\cosh2\phi_1],\nonumber \\
& &\hat{\xi}^{15}=-\hat{\xi}^{24}=\frac{1}{2}(g_2-g_1)\sinh2\phi_1\sinh\phi_3,\nonumber \\
& &\hat{\xi}^{45}=\frac{1}{2}[g_1+g_2+(g_2-g_1)\cosh2\phi_1]\sinh^2\phi_3\label{hat_xi_def1}
\end{eqnarray}
and
\begin{eqnarray}
& &\hat{f}_{345}=-h_1\cosh^2\phi_3,\nonumber \\
& &\tilde{f}^{135}=-\tilde{f}^{153}=-\tilde{f}^{234}=\tilde{f}^{243}=-\frac{1}{2}h_1\sinh2\phi_3
\end{eqnarray}
as well as the vielbein on the $SO(5,2)/SO(5)\times SO(2)$ scalar coset
\begin{eqnarray}
& &{P^1}_m=-\frac{i}{2}(\cosh\phi_3\phi_1',0,0,\phi_3',0),\nonumber \\
& &{P^2}_m=-\frac{i}{2}(0,\cosh\phi_3\phi_1',0,0,\phi_3').
\end{eqnarray}       
\indent We can now perform the analysis of the supersymmetry transformations of $\psi_{\mu i}$, $\chi_i$ and $\lambda^a_i$ to obtain the relevant BPS equations. The detailed analysis is given in the appendix. The resulting BPS equations from $\delta\lambda^a_i=0$ read 
\begin{eqnarray}
\frac{1}{2}\left(\cosh\phi_3\phi'_1-\frac{1}{\sqrt{2}}\Sigma^2\tilde{\xi}^{12}\right)\cos\theta+\frac{1}{4}\Sigma e^{-2h}\coth\phi_3K_4\sin\theta&=&0,\nonumber \\
\frac{1}{2}\left(\cosh\phi_3\phi'_1+\frac{1}{\sqrt{2}}\Sigma^2\tilde{\xi}^{12}\right)\sin\theta+\frac{1}{4}\Sigma e^{-2h}\coth\phi_3K_4\cos\theta&=&0,\nonumber \\
\frac{1}{2}\left(\phi_3'-\Sigma^{-1}\tilde{f}^{135}-\frac{1}{\sqrt{2}}\Sigma^2\tilde{\xi}^{15}\right)\cos\theta&=&0,\nonumber \\
-\frac{1}{2}\left(\phi_3'+\Sigma^{-1}\tilde{f}^{135}-\frac{1}{\sqrt{2}}\Sigma^2\tilde{\xi}^{15}\right)\sin\theta&=&0\label{dlambda1}
\end{eqnarray}
and
\begin{eqnarray}
-\frac{1}{2}\left(\cosh\phi_3\phi'_1-\frac{1}{\sqrt{2}}\Sigma^2\tilde{\xi}^{12}\right)\cos\theta+\frac{1}{4}\Sigma e^{-2f}\coth\phi_3K_5\sin\theta&=&0,\nonumber \\
\frac{1}{2}\left(\cosh\phi_3\phi'_1+\frac{1}{\sqrt{2}}\Sigma^2\tilde{\xi}^{12}\right)\sin\theta+\frac{1}{4}\Sigma e^{-2f}\coth\phi_3K_5\cos\theta&=&0,\nonumber \\
-\frac{1}{2}\left(\phi_3'-\Sigma^{-1}\tilde{f}^{135}-\frac{1}{\sqrt{2}}\Sigma^2\tilde{\xi}^{15}\right)\cos\theta&=&0,\nonumber \\
\frac{1}{2}\left(\phi_3'+\Sigma^{-1}\tilde{f}^{135}-\frac{1}{\sqrt{2}}\Sigma^2\tilde{\xi}^{15}\right)\sin\theta&=&0\label{dlambda2}
\end{eqnarray}
with
\begin{eqnarray}
& &K_4=(g_2\tilde{b}_7\cosh\phi_1-g_1\tilde{b}_2\sinh\phi_1)\sinh\phi_3,\nonumber \\
\textrm{and}\qquad & & K_5=(g_1b_1\sinh\phi_1-g_2b_6\cosh\phi_1)\sinh\phi_3\, .\label{K4_K5_def}
\end{eqnarray}
From $\delta\chi_i=0$, we find
\begin{eqnarray}
\frac{1}{2\sqrt{3}}\left(3\frac{\Sigma'}{\Sigma}+\Sigma^{-1}\hat{f}_{345}-\sqrt{2}\Sigma^2(\hat{\xi}^{12}+\hat{\xi}^{45})\right)\cos\theta\qquad\qquad & & \nonumber \\
+\frac{1}{4\sqrt{3}}\Sigma\left(e^{-2f}K_2-e^{-2h}K_1\right)\sin\theta&=&0,\nonumber \\
\frac{1}{2\sqrt{3}}\left(3\frac{\Sigma'}{\Sigma}+\Sigma^{-1}\hat{f}_{345}+\sqrt{2}\Sigma^2(\hat{\xi}^{12}+\hat{\xi}^{45})\right)\sin\theta\qquad\qquad & & \nonumber \\-\frac{1}{4\sqrt{3}}\Sigma\left(e^{-2f}K_2+e^{-2h}K_1\right)\cos\theta&=&0,\nonumber \\
\left(e^{-2h}K_4+e^{-2f}K_5\right)\sin\theta=\left(e^{-2h}K_4+e^{-2f}K_5\right)\cos\theta&=&0\, .\label{dchi}
\end{eqnarray}
with
\begin{eqnarray}
& &K_1=g_2\tilde{b}_7\sinh\phi_1-g_1\tilde{b}_2\cosh\phi_1,\nonumber \\
\textrm{and}\qquad & & K_2=g_1b_1\cosh\phi_1-g_2b_6\sinh\phi_1\, .\label{K1_K2_def}
\end{eqnarray}
The variation of the gravitinos along $AdS_2$ gives
\begin{eqnarray}
\left\{-\frac{1}{2}\left[f'+\frac{1}{3\sqrt{2}}\Sigma^2(\hat{\xi}^{12}+\hat{\xi}^{45})+\frac{1}{3}\Sigma^{-1}\hat{f}_{345}\right]\cos\theta\right. \quad & & \nonumber \\
\left.-\frac{\kappa}{2}e^{-f}e^{-2i\varphi\gamma_{\hat{r}}}\sin\theta \Gamma_1-\frac{1}{12}\Sigma(K_1e^{-2h}+2K_2e^{-2f})\sin\theta\right\}\tilde{\epsilon}&=&0, \nonumber \\
\left\{-\frac{1}{2}\left[f'-\frac{1}{3\sqrt{2}}\Sigma^2(\hat{\xi}^{12}+\hat{\xi}^{45})+\frac{1}{3}\Sigma^{-1}\hat{f}_{345}\right]\sin\theta\right. \quad & & \nonumber \\
\left.+\frac{\kappa}{2}e^{-f}e^{-2i\varphi\gamma_{\hat{r}}}\cos\theta \Gamma_1-\frac{1}{12}\Sigma(K_1e^{-2h}-2K_2e^{-2f})\cos\theta\right\}\tilde{\epsilon}&=&0, \nonumber \\
\left(e^{-2h}K_4-2e^{-2f}K_5\right)\sin\theta=\left(e^{-2h}K_4-2e^{-2f}K_5\right)\cos\theta &=&0\label{dpsiAdS2}
\end{eqnarray}
in which we have omitted the indices $i,j$ for convenience. Similary, the variation $\delta \psi_{s i}$ gives
\begin{eqnarray}
\left\{\frac{1}{2}\left[h'+\frac{1}{3\sqrt{2}}\Sigma^2(\hat{\xi}^{12}+\hat{\xi}^{45})+\frac{1}{3}\Sigma^{-1}\hat{f}_{345}\right]\cos\theta\right. \quad & & \nonumber \\
\left.-\frac{\tilde{\kappa}}{2}e^{-h}e^{-2i\varphi\gamma_{\hat{r}}}\sin\theta \Gamma_1+\frac{1}{12}\Sigma(2K_1e^{-2h}+K_2e^{-2f})\sin\theta\right\}\tilde{\epsilon}&=&0, \nonumber \\
\left\{\frac{1}{2}\left[-h'+\frac{1}{3\sqrt{2}}\Sigma^2(\hat{\xi}^{12}-\hat{\xi}^{45})+\frac{1}{3}\Sigma^{-1}\hat{f}_{345}\right]\sin\theta\right. \quad & & \nonumber \\
\left.-\frac{\tilde{\kappa}}{2}e^{-h}e^{-2i\varphi\gamma_{\hat{r}}}\cos\theta \Gamma_1+\frac{1}{12}\Sigma(2K_1e^{-2h}-K_2e^{-2f})\cos\theta\right\}\tilde{\epsilon}&=&0, \nonumber \\
\left(e^{-2f}K_5-2e^{-2h}K_4\right)\sin\theta=\left(2e^{-2h}K_4-e^{-2f}K_5\right)\cos\theta &=&0\, .\label{dpsiS2}
\end{eqnarray}
Finally, $\delta\psi_{ri}$ leads to
\begin{eqnarray}
\left[\frac{Y'}{Y}+\frac{1}{6\sqrt{2}}\Sigma^2(\hat{\xi}^{12}+\hat{\xi}^{45})+\frac{1}{6}\Sigma^{-1}\hat{f}_{345}\right]\cos\theta & &\nonumber \\
-\left[\theta'+\frac{1}{12}\Sigma(e^{-2h}K_1-e^{-2f}K_2)\right]\sin\theta&=&0,\nonumber \\
\left[\frac{Y'}{Y}+\frac{1}{6\sqrt{2}}\Sigma^2(\hat{\xi}^{12}+\hat{\xi}^{45})-\frac{1}{6}\Sigma^{-1}\hat{f}_{345}\right]\sin\theta & &\nonumber \\
+\left[\theta'-\frac{1}{12}\Sigma(e^{-2h}K_1+e^{-2f}K_2)\right]\cos\theta&=&0,\nonumber \\
\varphi'\cos\theta=\varphi'\sin\theta&=&0,\nonumber \\
\left(e^{-2h}K_4+e^{-2f}K_5\right)\sin\theta=\left(e^{-2h}K_4+e^{-2f}K_5\right)\cos\theta&=&0\, . \label{dpsir}
\end{eqnarray}
In deriving these equations, we have imposed two projectors
\begin{equation}
\gamma_{\hat{r}}\tilde{\epsilon}_i=i\Gamma_{12}\tilde{\epsilon}_i\qquad \textrm{and}\qquad \Gamma_3\tilde{\epsilon}_i=-\tilde{\epsilon}_i\, .\label{proj}
\end{equation}
From the BPS equations, we find that the last two equations in \eqref{dlambda1} and \eqref{dlambda2} imply either $\phi_3=0$ or $\theta=0,\frac{\pi}{2}$. Furthermore, compatibility between the first two equations in \eqref{dlambda1} and \eqref{dlambda2} requires $K_4\coth\phi_3=K_5\coth\phi_3=0$ which implies 
\begin{equation}
g_2\tilde{b}_7\cosh\phi_1=g_1\tilde{b}_2\sinh\phi_1\qquad \textrm{and}\qquad g_1b_1\sinh\phi_1=g_2b_6\cosh\phi_1\, .\label{K4_K5_con2}
\end{equation}
However, it turns out that these two algebraic constraints are not compatible with the remaining BPS equations and the two-form field equations \eqref{2-form_eq} unless $\phi_1=0$. With $\phi_1=0$, the two conditions in \eqref{K4_K5_con2} further require that
\begin{equation}
b_6=\tilde{b}_7=0
\end{equation}
for $g_2\neq0$ or $g_2=0$ with $b_6\neq0$ and $\tilde{b}_7\neq 0$. However, for the second possibility with $\phi_1=0$ and $g_2=0$, it turns out that the two-fom field equations give $b_6'=\tilde{b}_7'=0$. Moreover, the $N=2$ supersymmetric $AdS_5$ vacuum disappears when $g_2=0$. Therefore, we will choose the constants $b_6$ and $\tilde{b}_7$ to zero and keep $g_2\neq 0$.
\\
\indent Since $\phi_1=0$, we need $\phi_3\neq0$ and $\theta=0,\frac{\pi}{2}$ in order to have solutions with non-vanishing scalars from vector multiplets. For $\theta=0$, the Killing spinors in \eqref{Killing_spin} simply become
\begin{equation}
\epsilon_i=Y\tilde{\epsilon}_i
\end{equation}
while setting $\theta=\frac{\pi}{2}$ gives the Killing spinors of the form
\begin{equation}
\epsilon_i=Y\gamma_{01}{(\Gamma_1)_i}^j\tilde{\epsilon}_j\, .
\end{equation}
Both choices lead to the same BPS equations upon flipping the sign of the $\gamma_r$-projector given in \eqref{proj} as well as redefining $\phi_3$ and $h_1$ to $-\phi_3$ and $-h_1$. For definiteness, we will choose $\theta=0$ and find the BPS equation for $\phi_3$ as follows 
\begin{equation}
\phi_3'-\Sigma^{-1}\tilde{f}^{135}-\frac{1}{\sqrt{2}}\Sigma^2\tilde{\xi}^{15}=0\, .
\end{equation}
Setting $\theta=0$ in the first two equations in \eqref{dchi} gives the BPS equation for $\Sigma$ 
\begin{eqnarray}
\frac{\Sigma'}{\Sigma}=-\frac{1}{3}\Sigma^{-1}\hat{f}_{345}+\frac{\sqrt{2}}{3}\Sigma^2(\hat{\xi}^{12}+\hat{\xi}^{45})
\end{eqnarray}
together with an algebraic constraint
\begin{equation}
e^{-2h}\tilde{b}_2=e^{-2f}b_1\, .\label{chi_con}
\end{equation}
In addition, with $\theta=0$, the two equations involving $\varphi'$ in \eqref{dpsir} give $\varphi'=0$. 
\\
\indent All the remaining equations then lead to
\begin{eqnarray}
f'&=&h'=\frac{2Y'}{Y}=-\frac{1}{3\sqrt{2}}\Sigma^2(\hat{\xi}^{12}+\hat{\xi}^{34})-\frac{1}{3}\Sigma^{-1}\hat{f}_{345},\nonumber \\
0&=&\left[\kappa e^{-2i\varphi\gamma_{\hat{r}}}\Gamma_1+\frac{1}{2}g_1\Sigma e^{-f}b_1\right]\tilde{\epsilon}=\left[\tilde{\kappa} e^{-2i\varphi\gamma_{\hat{r}}}\Gamma_1+\frac{1}{2}g_1\Sigma e^{-f}\tilde{b}_2\right]\tilde{\epsilon}\, .\label{phi10_eq1}
\end{eqnarray}
From these equations, we find $f=h$ up to an irrelevant integration constant. Equation \eqref{chi_con} then implies that $\tilde{b}_2=b_1$. Therefore, the only non-vanishing two-form fields take the form 
\begin{equation}
B_1=b_1\textrm{vol}_{AdS_2} \qquad \textrm{and}\qquad B_2=b_1\textrm{vol}_{S^2}\, .
\end{equation}
\indent As in \cite{line_defects_5DN4_Nicolo}, we have chosen the basis for $\Gamma_m$ such that $\Gamma_1$ is diagonal. In this case, we can solve the last two equations in \eqref{phi10_eq1} by choosing $\varphi=0$ and $\kappa_i=\tilde{\kappa}_i$ with
\begin{equation}
\kappa_1=\kappa_4=-\kappa_2=-\kappa_3=-\kappa=\pm 1
\end{equation}
leading to the BPS equation for $b_1$ of the form
\begin{equation}
b_1=-\frac{2\kappa}{g_1\Sigma}e^f\, .
\end{equation}
With all these, the two-form field equations in \eqref{2-form_eq} reduce to a single equation
\begin{equation}
b_1'-\frac{1}{\sqrt{2}}g_1b_1\Sigma^2=0\, .
\end{equation}
Consistency between this equation and the previously derived BPS equations implies
\begin{equation}
g_2\Sigma^2\sinh^2\phi_3=0\, .
\end{equation}
For $\phi_3=0$, all the scalars from vector multiplets vanish leading to solutions of pure $N=4$ gauged supergravity. To find solutions of the matter-coupled gauged supergravity, we then set $g_2=0$. This in turn implies that the $N=2$ supersymmetric $AdS_5$ vacuum does not appear in the BPS equations, so there is no solution that is asymptotic to this $AdS_5$ geometry. We end this section by noting that for $\phi_1=0$, the $\Gamma_3$ projector in \eqref{proj} is not needed, see more detail in the appendix, so the solutions preserve half of the original supersymmetry. 
 
\subsection{Line defect solutions}
We now look for possible solutions to the BPS equations derived in the previous section. Before solving the BPS equations, we first note that the resulting BPS equations for scalar fields are essentially the same as those of holographic RG flows studied in \cite{5D_N4_flow_Davide,5D_flowII}. In addition, the contribution of the two-form fields to the scalar field equations, with the present ansatz for various fields, comes from the following terms in the Lagrangian
\begin{equation}
g_1^2M_{11}(\tilde{b}_2^2e^{4h}-b_1^2e^{-4f})+g_2^2M_{66}(\tilde{b}_7^2e^{-4h}-b_6^2e^{-4f})-2g_1g_2M_{16}(\tilde{b}_2\tilde{b}_7e^{-4h}-b_1b_6e^{-4f}).\label{2-form_con}
\end{equation}
In all cases considered in this paper, we have $f=h$ and $\tilde{b}_2=b_1$ and $b_6=\tilde{b}_7=0$ for which all the terms in \eqref{2-form_con} vanish. Therefore, all scalar field equations are satisfied as in the case of holographic RG flows studied in \cite{5D_N4_flow_Davide,5D_flowII}. Furthermore, in all cases, the function $Y$ satisfies the relation $Y'=\frac{Y}{2}f'$ which gives 
\begin{equation}
Y=Y_0e^{\frac{f}{2}}\, .
\end{equation}
The integration constant $Y_0$ can also be set to unity. Since both the warp factors of $AdS_2$ and $S^2$ are equal, the solutions are called charged domain walls in \cite{line_defects_5DN4_Nicolo}.
\\ 
\indent Setting $\phi_1=0$ and $g_2=0$, we find that the BPS equations are explicitly given by
\begin{eqnarray}
\phi_3'&=&-\frac{1}{2}h_1\Sigma^{-1}\sinh2\phi_3,\label{phi3_eq1} \\ 
\Sigma'&=&\frac{1}{3}h_1\cosh^2\phi_3-\frac{\sqrt{2}}{3}g_1\Sigma^3,\\
f'&=&\frac{1}{3}h_1\Sigma^{-1}\cosh^2\phi_3+\frac{1}{3\sqrt{2}}g_1\Sigma^2
\end{eqnarray}
together with 
\begin{equation}
b_1=-\frac{2\kappa}{g_1\Sigma}e^f\, .
\end{equation}
From these equations, we see that there is a supersymmetric $AdS_5$ vacuum corresponding to a fixed point solution of $\phi_3'=\Sigma'=0$ and $f'=\frac{1}{L}$ with $L$ being the $AdS_5$ radius. This $AdS_5$ vacuum preserves the full $N=4$ supersymmetry and is given by
\begin{equation}
\phi_3=0,\qquad \Sigma=\left(\frac{h_1}{\sqrt{2}g_1}\right)^{\frac{1}{3}},\qquad L^2=\frac{2}{h_1}\left(\frac{h_1}{\sqrt{2}g_1}\right)^{\frac{1}{3}}\, .\label{AdS5_1}
\end{equation}
We also note that the value of the dilaton $\Sigma$ at the vacuum can be shifted such that $\Sigma=1$ at the $AdS_5$ vacuum. This is achieved by setting $h_1=\sqrt{2}g_1$. 
\\
\indent Unlike the case of RG flow solutions which can be obtained only numerically, the vanishing of $g_2$ allows to analytically find the solutions. Moreover, in the absence of two-form fields, the solutions would describe holographic RG flows from the $N=2$ SCFT dual to the $AdS_5$ vacuum in \eqref{AdS5_1} to non-conformal phases in the IR corresponding to singular geometries. For $SO(2)_D\times SO(3)$ gauged supergravity, this type of solutions has not been previously considered in \cite{5D_N4_flow_Davide,5D_flowII}, so we will give some detail for obtaining the solutions. We proceed by first consider a linear combination
\begin{equation}
2f'+\frac{\Sigma'}{\Sigma}=h_1\Sigma^{-1}\cosh^2\phi_3\, .
\end{equation}
By taking $\phi_3$ as an independent variable instead of $r$ and using the $\phi_3'$ equation in \eqref{phi3_eq1}, we can solve for $f(\phi_3)$ of the form
\begin{equation}
f=-\frac{1}{2}\ln \sinh\phi_3-\frac{1}{2}\ln \Sigma\, .
\end{equation}
We have omitted an additive integration constant to $f$ since this can be absorbed by suitable rescaling of coordinates.
\\
\indent Similarly, we can find the solution for $\Sigma(\phi_3)$ given by
\begin{equation}
\Sigma^3=\frac{h_1}{\sqrt{2}g_1+\left(C_0h_1+2\sqrt{2}g_1\tan^{-1}\tanh\frac{\phi_3}{2}\right)\sinh\phi_3}
\end{equation}
with $C_0$ being an integration constant. We also note that the solution approach the $AdS_5$ vacuum \eqref{AdS5_1} for any values of $C_0$.
\\
\indent With all these solutions and a new radial coordinate $\tilde{r}$ defined by $\frac{d\tilde{r}}{dr}=\Sigma^{-1}$, we find the solution for $\phi_3$ given by
\begin{equation}
h_1(\tilde{r}-\tilde{r}_0)=\ln\cosh\phi_3-\ln \sinh\phi_3\, .
\end{equation}
The integration constant $\tilde{r}_0$ can also be set to zero by shifting the coordinate $\tilde{r}$. Finally, we can use all these results to determine the non-vanishing two-form fields
\begin{equation}
B_1=b_1\textrm{vol}_{AdS_2}\qquad \textrm{and}\qquad B_2=b_1\textrm{vol}_{S^2}
\end{equation}
with
\begin{equation}
b_1=-\frac{2\kappa}{g_1}\sqrt{\frac{h_1C_0+2\sqrt{2}g_1\tan^{-1}\tanh\frac{\phi_3}{2}+\sqrt{2}g_1\textrm{csch}\phi_3}{h_1g_1^2}}\, .
\end{equation} 
\indent As $\phi_3\sim 0$, the solution is asymptotic to the $AdS_5$ vacuum \eqref{AdS5_1} with
\begin{eqnarray}
\phi_3\sim e^{-\frac{2r}{L}},\quad \Sigma\sim \left(\frac{h_1}{\sqrt{2}g_1}\right)^{\frac{1}{3}}\left(1-\frac{C_0h_1}{3\sqrt{2}g_1} e^{-\frac{2r}{L}}\right),\quad f\sim \frac{r}{L},\quad b_1\sim e^{\frac{r}{L}}\, .
\end{eqnarray}
We also note that, in this limit, $\tilde{r}\sim r$. From this behavior, we see that both $\phi_3$ and $\Sigma$ correspond to vacuum expectation values of operators of dimension $\Delta=2$. The constant $C_0$ is related to the vacuum expectation value of the operator dual to $\Sigma$. By linearizing the two form field equations in \eqref{2-form_eq} as in \cite{line_defects_5DN4_Nicolo}, we find that $b_1$ is dual to an operator of dimension $\Delta=3$. Therefore, in addition to the vacuum expectation values in the case of RG flows, the solution contains a source term for dimension-3 operators dual to the two-form fields $B_1$ and $B_2$. Due to the non-vanishing two-forms, the asymptotic geometry is only locally $AdS_5$ with the contribution from the two-form fields being subleading to the scalar potential as pointed out in \cite{line_defects_5DN4_Nicolo}.
\\
\indent As $\tilde{r}\rightarrow \tilde{r}_0$, the solution is singular with, for $C_0\neq -\frac{g_1\pi}{\sqrt{2}h_1}$, 
\begin{eqnarray}
& &\phi_3\sim -\frac{1}{2}\ln (\tilde{r}-\tilde{r}_0),\qquad \Sigma\sim (\tilde{r}-\tilde{r}_0)^{\frac{1}{6}},\nonumber \\
& &f\sim \frac{1}{6}\ln (\tilde{r}-\tilde{r}_0),\qquad b_1\sim \textrm{constant}\, .
\end{eqnarray}
In this limit, the metric becomes
\begin{equation}
ds^2=(\tilde{r}-\tilde{r}_0)^{\frac{1}{3}}(ds^2_{AdS_2}+ds^2_{S^2}+d\tilde{r}^2).
\end{equation}
To determine whether this singularity is physically acceptable or not, we check the criterion given in \cite{Gubser_Sing}. The scalar potential with $g_2=0$ and only $\phi_3$ and $\Sigma$ non-vanishing is given by
\begin{equation}
V=\frac{1}{4}\Sigma^{-2}\cosh^2\phi_3\left[h_1^2(\cosh2\phi_3-3)-4\sqrt{2}g_1h_1\Sigma^3\right].
\end{equation} 
Near the singularity, we find that $V\rightarrow +\infty$, so the singularity is not acceptable by the criterion of \cite{Gubser_Sing}.
\\
\indent For $C_0=-\frac{g_1\pi}{\sqrt{2}h_1}$, we find the asymptotic behavior near $\tilde{r}_0$ as
\begin{eqnarray}
& &\phi_3\sim -\frac{1}{2}\ln (\tilde{r}-\tilde{r}_0),\qquad \Sigma\sim (\tilde{r}-\tilde{r}_0)^{-\frac{1}{3}},\nonumber \\
& &f\sim \frac{5}{6}\ln (\tilde{r}-\tilde{r}_0),\qquad b_1\sim (\tilde{r}-\tilde{r}_0)^{\frac{7}{6}}
\end{eqnarray}
which gives $V\rightarrow -\infty$. Therefore, in this case, the singularity is acceptable with the five-dimensional metric near the singularity given by
\begin{equation}
ds^2=(\tilde{r}-\tilde{r}_0)^{\frac{5}{3}}\left[ds^2_{AdS_2}+ds^2_{S^2}+(\tilde{r}-\tilde{r}_0)^{-\frac{7}{3}}d\tilde{r}^2\right].
\end{equation}
\indent Although the $SO(2)_D\times SO(3)$ gauged supergravity can be embedded in eleven-dimensional supergravity, the complete truncation ansatz has not been worked out to date, so we currently cannot determine whether the uplifted solutions to eleven dimensions are physical or not.  

\section{Line defects from $SO(2)\times ISO(3)$ gauged supergravity}\label{SO2_ISO3}
In this section, we consider line defect solutions in $N=4$ gauged supergravity coupled to three vector multiplets with $SO(2)\times ISO(3)\sim SO(2)\times (SO(3)\ltimes \mathbb{R}^3)$ gauge group. The embedding tensor is given by
\begin{eqnarray}
\xi^{\hat{m}\hat{n}}&=&g_1\epsilon_{\hat{m}\hat{n}},\qquad \hat{m},\hat{n}=1,2,\nonumber \\  
f_{\tilde{m}\tilde{n}\tilde{p}}&=&g\epsilon_{\tilde{m}\tilde{n}\tilde{p}},\qquad \tilde{m},\tilde{n},\tilde{p}=3,4,5,\nonumber \\
f_{a+5,b+5,c+5}&=&-2g\epsilon_{abc},\qquad f_{a+2,b+5,c+5}=-g\epsilon_{abc}, \qquad a,b,c=1,2,3
\end{eqnarray} 
with the gauge coupling constants $g_1$ and $g$. This gauged supergravity admits an $N=4$ supersymmetric $AdS_5$ vacuum at the origin of $SO(5,3)/SO(5)\times SO(3)$ with 
\begin{equation}
\Sigma=-\left(\frac{g}{\sqrt{2}g_1}\right)^{\frac{1}{3}},\qquad V_0=-3\left(\frac{g^2g_1}{2}\right)^{\frac{2}{3}},\qquad L=\left(-\frac{4\sqrt{2}}{g^2g_1}\right)^{\frac{1}{3}}\, .
\end{equation}
We can also choose $g=-\sqrt{2}g_1$ in order to have $\Sigma=1$ at the $AdS_5$ vacuum. This $AdS_5$ vacuum can be identified as an $AdS_5\times H^2\times S^4$ solution of eleven-dimensional supergravity \cite{Malek_AdS5_N4_embed,ISO3_5D_N4_gauntlett}. A number of holographic solutions with vanishing two-form fields in this gauged supergravity have been studied recently in \cite{Janus_5D_ISO3}. In this paper, we will look for solutions with non-vanishing two-form fields describing conformal line defects within the $N=2$ SCFT dual to the supersymmetric $AdS_5$ vacuum. This SCFT arises from $N=(0,2)$ SCFT in six dimensions compactified on $H^2$.
\\
\indent We will consider solutions with $SO(2)_{\textrm{diag}}$ symmetry. As shown in \cite{Janus_5D_ISO3}, there are five singlet scalars with the corresponding coset representative given by 
\begin{equation}
\mc{V}=e^{\phi_1(Y_{12}+Y_{23})}e^{\phi_2(Y_{13}-Y_{22})}e^{\phi_3(Y_{42}+Y_{53})}e^{\phi_4(Y_{43}-Y_{52})}e^{\phi_5Y_{31}}\, .\label{ISO3_SO2d_coset}
\end{equation}
As in the previous case, to consistently truncate out all the vector fields, we need to set $\phi_2=\phi_4=0$. With this, we find the vielbein on the scalar manifold
\begin{eqnarray}
{P^1}_m&=&-\frac{i}{2}(0,0,\phi'_5,0,0),\nonumber \\
{P^2}_m&=&-\frac{i}{2}(\cosh\phi_3\phi'_1,0,0,\phi_3',0),\nonumber \\
{P^3}_m&=&-\frac{i}{2}(0,\cosh\phi_3\phi_1',0,0,\phi_3').
\end{eqnarray}
All non-vanishing dressed components of the embedding tensor read
\begin{eqnarray}
\tilde{f}^{112}&=&g\sinh^2\phi_1(\sinh\phi_5-2\cosh\phi_5),\nonumber \\
\tilde{f}^{115}&=&-\tilde{f}^{124}\nonumber \\
&=&g\sinh\phi_1\left[\cosh\phi_3\cosh\phi_5+\cosh\phi_1\sinh\phi_3(\sinh\phi_5-2\cosh\phi_5)\right],\nonumber \\
\tilde{f}^{145}&=&-g(\cosh\phi_3-\cosh\phi_1\sinh\phi_3)\left[\cosh\phi_3\sinh\phi_5\right.\nonumber \\
& &\left.+\cosh\phi_1\sinh\phi_3(\sinh\phi_5-2\cosh\phi_5)\right],\nonumber \\
\tilde{f}^{223}&=&-\tilde{f}^{313}\nonumber \\
&=&g\sinh\phi_1\left[\cosh\phi_1\cosh\phi_3(\cosh\phi_3-2\sinh\phi_5)+\sinh\phi_3\sinh\phi_5\right],\nonumber 
\end{eqnarray}
\begin{eqnarray}
\tilde{f}^{235}&=&-\tilde{f}^{334}\nonumber \\
&=&-\frac{1}{2}g\left[\cosh\phi_5\sinh^2\phi_1\sinh2\phi_3\right.\nonumber \\
& &\left.+2\cosh\phi_1\sinh\phi_5(\cosh2\phi_3-\cosh\phi_1\sinh2\phi_3)\right],
\end{eqnarray}
\begin{eqnarray}
\tilde{\xi}^{2m}&=&\frac{1}{2}g_1(0,\cosh\phi_3\sinh2\phi_1,0,0,\sinh^2\phi_1\sinh2\phi_3),\nonumber \\
\tilde{\xi}^{3m}&=&-\frac{1}{2}g_1(\cosh\phi_3\sinh2\phi_1,0,0,\sinh^2\phi_1\sinh2\phi_3,0),\nonumber \\
\hat{\xi}^{12}&=&g_1\cosh^2\phi_1,\nonumber \\
\hat{\xi}^{15}&=&-\hat{\xi}^{24}=\frac{1}{2}g_1\sinh2\phi_1\sinh\phi_3,\nonumber \\
\hat{\xi}^{45}&=&g_1\sinh^2\phi_1\sinh^2\phi_3,
\end{eqnarray}
and
\begin{eqnarray}
\hat{f}_{123}&=&-g\sinh^2\phi_1(\cosh\phi_5-2\sinh\phi_5),\nonumber \\
\hat{f}_{135}&=&-\hat{f}_{234}\nonumber \\
&=&g\sinh\phi_1\left[\cosh\phi_1\sinh\phi_3(\cosh\phi_5-2\sinh\phi_5)+\cosh\phi_5\sinh\phi_5\right],\nonumber \\
\hat{f}_{345}&=&g(\cosh\phi_3-\cosh\phi_1\sinh\phi_3)\left[\cosh\phi_3\cosh\phi_5\right.\nonumber \\
& &\left.+\cosh\phi_1\sinh\phi_3(\cosh\phi_5-2\sinh\phi_5)\right].
\end{eqnarray}
\indent Among eight vector fields $(A^m_\mu,A^a_\mu)$, only $A^1_\mu$ and $A^2_\mu$ are charged under the $SO(2)$ factor of the gauge group which is gauged by $A^0_\mu$ in the supergravity multiplet. Therefore, in this case, there are only two two-form fields $B_{1\mu\nu}$ and $B_{2\mu\nu}$ arising from dualizing $A^1_\mu$ and $A^2_\mu$ to implement the gauging. We can now carry out the same analysis as in the previous section. In particular, we will use the same ansatze for the metric, two-form fields and Killing spinors with an obvious modification $B_6=B_7=0$. However, the same analysis of the two-form field equations is still valid and leads to 
\begin{equation}
B_1=b_1\textrm{vol}_{AdS_2}\qquad \textrm{and}\qquad B_2=\tilde{b}_2\textrm{vol}_{S^2}\, .
\end{equation}
With all these, we find that the supersymmetry transformation $\delta\lambda^1_i$ gives
\begin{equation}
\phi'_5=\Sigma^{-1}(\tilde{f}^{112}+\tilde{f}^{145}).
\end{equation}
From $\delta\lambda^2_i$ and $\delta\lambda^3_i$, we find two sets of equations
\begin{eqnarray}
\left(\cosh\phi_3\phi_1'-\Sigma^{-1}\tilde{f}^{223}+\frac{1}{\sqrt{2}}\Sigma^2\tilde{\xi}^{22}\right)\cos\theta&=&0,\nonumber\\ 
\left(\cosh\phi_3\phi_1'-\Sigma^{-1}\tilde{f}^{223}-\frac{1}{\sqrt{2}}\Sigma^2\tilde{\xi}^{22}\right)\sin\theta&=&0\label{phi1_eq22}
\end{eqnarray}
and
\begin{eqnarray}
\left(\phi_3'+\Sigma^{-1}\tilde{f}^{235}+\frac{1}{\sqrt{2}}\Sigma^2\tilde{\xi}^{25}\right)\cos\theta&=&0,\nonumber \\
\left(\phi_3'-\Sigma^{-1}\tilde{f}^{235}+\frac{1}{\sqrt{2}}\Sigma^2\tilde{\xi}^{25}\right)\sin\theta&=&0\label{phi3_eq22}
\end{eqnarray}
together with two algebraic constraints
\begin{equation}
b_1\sinh\phi_1e^{-2f}=\tilde{b}_2\sinh\phi_1e^{-2h}=0\, .
\end{equation}
Unlike in the previous section, the last two equations imply that $\phi_1=0$ for non-vanishing two-form fields. Therefore, solutions with non-vanishing scalar fields from vector multiplets are possible only with $\phi_3\neq 0$ or $\phi_5\neq 0$. Since only $\phi_1$ is charged under the $SO(2)$ part of the gauge group, the symmetry of the solutions is enhanced to $SO(2)\times SO(2)$ with the second $SO(2)$ being a subgroup of $SO(3)\subset ISO(3)$. It can also be verified that setting $\phi_1=0$ is consistent with \eqref{phi1_eq22}. Equation \eqref{phi3_eq22} then implies that either $\theta=0$ or $\theta=\frac{\pi}{2}$ for $\phi_3\neq 0$. As in the previous case, the two choices of $\theta=0$ and $\theta=\frac{\pi}{2}$ lead to equivalent BPS equations after flipping the sign of $\gamma_r$-projector and a suitable redefinition of some fields and parameters. To simplify subsequent results, we will from now on set
\begin{equation}
\phi_1=0\qquad \textrm{and}\qquad \theta=0\, .
\end{equation}
\indent $\delta \chi_i=0$ conditions then give
\begin{equation}
\frac{\Sigma'}{\Sigma}=-\frac{1}{3}\Sigma^{-1}\hat{f}_{345}+\frac{\sqrt{2}}{3}\Sigma^2\hat{\xi}^{12}
\end{equation} 
and
\begin{equation}
\tilde{b}_2e^{-2h}=b_1e^{-2f}\, .
\end{equation}
Finally, $\delta\psi_{\mu i}=0$ conditions lead to 
\begin{equation}
2\frac{Y'}{Y}=h'=f'=-\frac{1}{3\sqrt{2}}\Sigma^2\hat{\xi}^{12}-\frac{1}{3}\Sigma^{-1}\hat{f}_{345}
\end{equation}
and
\begin{equation}
\left(\kappa e^{-2i\varphi\gamma_{\hat{r}}}\Gamma_1-\frac{1}{2}g_1\Sigma b_1e^{-f}\right)\tilde{\epsilon}=0,\qquad \left(\tilde{\kappa} e^{-2i\varphi\gamma_{\hat{r}}}\Gamma_1-\frac{1}{2}g_1\Sigma \tilde{b}_2e^{-h}\right)\tilde{\epsilon}=0
\end{equation}
together with $\varphi'=0$. We again find that $h=f$ which leads to $\tilde{b}_2=b_1$. Choosing the constant $\varphi=0$ and $\kappa_i=\tilde{\kappa}_i$ with $\kappa_1=\kappa_4=-\kappa_2=-\kappa_3=\kappa$, we find that the last two equations reduce to
\begin{equation}
\kappa-\frac{1}{2}g_1\Sigma b_1e^{-f}=0\, .
\end{equation}
Setting $\phi_1=0$ and $\tilde{b}_2=b_1$, we are left with a single equation from the two-form field equations
\begin{equation}
b_1'-\frac{1}{\sqrt{2}}g_1\Sigma^2b_1=0\, .
\end{equation}
Unlike the previous case of $SO(2)_D\times SO(3)$ gauge group, this equation is automatically compatible with all the BPS equations without any additional constraints. The solutions also take the form of charged domain walls with the BPS equations are those of holographic RG flows with non-vanishing two-form fields. It can be readily verified that all the field equations are satisfied by the resulting BPS equations as in the previous section.    

\subsection{Solution with $SO(2)\times SO(3)$ symmetry}
We begin with a simpler solution with $SO(2)\times SO(3)$ symmetry by setting 
\begin{equation}
\phi_5=\phi_3\, .
\end{equation}
This is a consistent truncation as pointed out in \cite{Janus_5D_ISO3}. The BPS equations are given by
\begin{eqnarray}
\phi'_3&=&ge^{-2\phi_3}\sinh\phi_3,\nonumber \\
\Sigma'&=&-\frac{1}{6}\left[ge^{-3\phi_3}(1-3e^{2\phi_3})-2\sqrt{2}g_1\Sigma^3\right],\nonumber \\
f'&=&-\frac{1}{6}\left[g\Sigma^{-1}(e^{-3\phi_3}-3e^{-\phi_3})+\sqrt{2}g_1\Sigma^2\right],\nonumber \\
b_1&=&\frac{2\kappa}{g_1\Sigma}e^{f}\, . 
\end{eqnarray}
As in the previous section, we can solve for $f$, $\Sigma$ and $b_1$ as functions of $\phi_3$. The solutions for $f$, $\Sigma$ and $\phi_3$ have already been given in \cite{Janus_5D_ISO3}, so the solution for a line defect in this case is then given by
\begin{eqnarray}
\Sigma^3&=&-\frac{g}{\sqrt{2}g_1\cosh\phi_3},\nonumber \\
f&=&\frac{1}{3}\phi_3+\frac{1}{3}\ln(1-e^{2\phi_3})+\frac{1}{6}\ln\left[\sqrt{2}g_1(e^{4\phi_3}-1)\right],\nonumber \\
b_1&=&-\frac{\kappa 2^{\frac{11}{12}}}{g^{\frac{1}{3}}}\sqrt{\frac{e^{4\phi_3}-1}{g_1}},\nonumber \\
g(\tilde{r}-\tilde{r}_0)&=&2e^{\phi_3}-\ln(1+e^{\phi_3})+\ln(1-e^{\phi_3})
\end{eqnarray}
with a constant $\tilde{r}_0$ and $\tilde{r}$ defined by $\frac{d\tilde{r}}{dr}=\Sigma^{-1}$. Near the $AdS_5$ vacuum, we have, see more detail in \cite{Janus_5D_ISO3},
\begin{equation}
\phi_3\sim e^{\frac{2r}{L}}\qquad \textrm{and}\qquad \Sigma\sim 1+e^{-\frac{2r}{L}}
\end{equation}
which leads to 
\begin{equation}
b_1\sim \sqrt{\phi_3}\sim e^{\frac{r}{L}}\, .
\end{equation}
We then find that there is a source term for an operator of dimension $\Delta=6$ dual to $\phi_3$ as well as a deformation by an operator of dimension $\Delta=3$ dual to $b_1$. On the other hand, there is no source term for the dimension-2 operator dual to $\Sigma$. As pointed out in \cite{Janus_5D_ISO3}, in this case, the dual $N=2$ SCFT appear in the IR while the UV field theory is given by $N=(0,2)$ SCFT in six dimensions arising from M5-branes wrapped on $H^2$. 
\\
\indent Since the $SO(2)\times ISO(3)$ gauged supergravity can be obtained from a consistent truncation of eleven-dimensional supergravity on $H^2\times S^4$, we can uplift the solution to eleven dimensions using the ansatz given in \cite{ISO3_5D_N4_gauntlett} and \cite{M-theory_on_S4_1,M-theory_on_S4_2}. The eleven dimensional metric is given by
\begin{equation}
ds^2_{11}=\Delta^{\frac{1}{3}}ds^2_7+\frac{1}{m^2}\Delta^{-\frac{2}{3}}T^{-1}_{ab}D\mu^aD\mu^b,\qquad a,b=1,2,\ldots, 5
\end{equation} 
with $m$ being the gauge coupling constant in seven dimensions. $\mu^a$ are coordinates on $S^4$ satisfying $\mu^a\mu^a=1$ and $D\mu^a=d\mu^a+mA^{ab}_{(1)}\mu^b$. In the present solutions, all five-dimensional gauge fields vanish, so the only non-vanishing gauge field in seven dimensions is given by
\begin{equation}
A^{12}_{(1)}=\frac{1}{m}\omega_H
\end{equation}
with $\omega_H$ being the spin connection on $H^2$. For the metric on $H^2$ of the form
\begin{equation}
ds^2_{H^2}=\bar{e}^{\bar{a}}\bar{e}^{\bar{a}}=d\vartheta^2+\sinh^2\vartheta d\varphi^2,\qquad \bar{a}=1,2,
\end{equation}
we have
\begin{equation}
\omega_H=\cosh\vartheta d\varphi\, .
\end{equation}
We also note that 
\begin{equation}
d\omega_H=\textrm{vol}_{H^2}\, .
\end{equation}
\indent The warp factor is defined by
\begin{equation}
\Delta=T_{ab}\mu^a\mu^b
\end{equation}
in which $T_{ab}$ is a unimodular $SL(5)$ matrix. In the present case, $T_{ab}$ is diagonal, see more detail in \cite{Janus_5D_ISO3}. This implies that all terms of the form
\begin{equation}
\epsilon_{a_1a_2a_3a_4a_5}\mu^b\mu^cT^{a_1b}DT^{a_2c}\wedge D\mu^{a_3}\wedge D\mu^{a_4}\wedge D\mu^{a_5} 
\end{equation}
vanish identically. Therefore, the four-form field strength is given by
\begin{eqnarray}
G_{(4)}&=&\frac{\Delta^{-2}}{4!m^3}\epsilon_{a_1a_2a_3a_4a_5}\left[6m\Delta F^{a_1a_2}_{(2)}\wedge D\mu^{a_3}\wedge D\mu^{a_4}T^{a_5b}\mu^b\right.\nonumber \\
& &\left.\phantom{I^{a_1}_{(2)}}-U\mu^{a_1}D\mu^{a_2}\wedge D\mu^{a_3}\wedge D\mu^{a_4}\wedge D\mu^{a_5}\right]-T_{ab}*S^a_{(3)}\mu^b\nonumber \\
& &+\frac{1}{m}S^a_{(3)}\wedge D\mu^a\, .
\end{eqnarray}
In the solutions under consideration here, there are only two three-form fields given by
\begin{equation}
S^1_{(3)}=\frac{1}{\sqrt{2}}m(B_2\wedge \bar{e}^2-B_1\wedge \bar{e}^1)\quad \textrm{and}\quad S^2_{(3)}=-\frac{1}{\sqrt{2}}m(B_2\wedge \bar{e}^2+B_1\wedge \bar{e}^1).
\end{equation}
Recall that 
\begin{equation}
B_1=b_1\textrm{vol}_{AdS_2}\qquad \textrm{and}\qquad B_2=b_1\textrm{vol}_{S^2},
\end{equation}
we have
\begin{eqnarray}
S^1_{(3)}&=&\frac{1}{\sqrt{2}}mb_1(\textrm{vol}_{S^2}\wedge\bar{e}^2-\textrm{vol}_{AdS_2}\wedge \bar{e}^1),\nonumber \\
S^2_{(3)}&=&-\frac{1}{\sqrt{2}}mb_1(\textrm{vol}_{S^2}\wedge\bar{e}^1+\textrm{vol}_{AdS_2}\wedge \bar{e}^2).
\end{eqnarray}
\indent For $SO(2)\times SO(3)$ symmetric solutions, the $SL(5)$ matrix takes the form
\begin{equation}
T_{ab}=\textrm{diag}(e^{-6\lambda},e^{-6\lambda},e^{4\lambda},e^{4\lambda},e^{4\lambda}).
\end{equation}
Using the following $S^4$ coordinates 
\begin{equation}
\mu^a=(\cos\xi\cos\theta,\cos\xi\sin\theta,\sin\xi \hat{\mu}^{\hat{a}}),\qquad \hat{\mu}^{\hat{a}} \hat{\mu}^{\hat{a}}=1,
\end{equation}
for $\hat{a}=1,2,3$, and the seven-dimensional metric of the form
\begin{equation}
ds^2_7=e^{-4\phi}(e^{2f}ds^2_{AdS_2}+dr^2+e^{2h}ds^2_{S^2})+e^{6\phi}\bar{e}^{\bar{a}}\bar{e}^{\bar{a}},
\end{equation}
we find the eleven-dimensional metric and the four-form field strength given by
\begin{eqnarray}
ds^2_{11}&=&\Delta^{\frac{1}{3}}\left[e^{2f-4\phi}(ds^2_{AdS_2}+ds^2_{S^2})+e^{-4\phi}\Sigma^2d\tilde{r}^2+e^{6\phi}\bar{e}^{\bar{a}}\bar{e}^{\bar{a}}\right]\nonumber \\
& &+\frac{1}{m^2}\Delta^{-\frac{2}{3}}\left[(e^{6\lambda}\sin^2\xi+e^{-4\lambda}\sin^2\xi)d\xi^2\right.\nonumber \\
& &\left.+\cos^2\xi (d\theta-\omega_H)^2+\sin^2\xi d\hat{\mu}^{\hat{a}}d\hat{\mu}^{\hat{a}}\right]\label{uplifted_11Dmetric}
\end{eqnarray}
and
\begin{eqnarray}
G_{(4)}&=&-\frac{1}{m^3}\Delta^{-2}U\sin^2\xi\cos\xi \textrm{vol}_{\tilde{S}^2}\wedge d\theta\wedge d\xi+\frac{1}{m^3}\Delta^{-1}e^{4\lambda}\textrm{vol}_{H^2}\wedge \textrm{vol}_{\tilde{S}^2}\nonumber \\
& &-\frac{1}{\sqrt{2}}b_1e^{-6\lambda-2\phi}\Sigma \cos\xi\left(\textrm{vol}_{S^2}\wedge d\tilde{r}\wedge \bar{e}^{'2}-\textrm{vol}_{AdS_2}\wedge d\tilde{r}\wedge \bar{e}^{'1}\right)\nonumber \\
& &+\frac{1}{\sqrt{2}}b_1\sin\xi d\xi\wedge \left(\textrm{vol}_{S^2}\wedge \bar{e}^{'2}-\textrm{vol}_{AdS_2}\wedge \bar{e}^{'1}\right)\nonumber \\
& &+\frac{1}{\sqrt{2}}b_1\cos\xi d\theta\wedge \left(\textrm{vol}_{S^2}\wedge \bar{e}^{'1}+\textrm{vol}_{AdS_2}\wedge \bar{e}^{'2}\right).\label{G4_1}
\end{eqnarray}
In these equations, the scalars $\phi$ and $\lambda$ are defined by
\begin{equation}
\phi=\frac{1}{10}(3\phi_3-\ln \Sigma),\qquad \lambda=-\frac{1}{10}(\phi_3+3\ln\Sigma)
\end{equation}
and 
\begin{eqnarray}
& &\Delta=e^{-6\lambda}\cos^2\xi+e^{4\lambda}\sin^2\xi,\nonumber \\
& &U=2(e^{-6\lambda}\cos^2\xi+e^{4\lambda}\sin^2\xi)-\Delta(2e^{-6\lambda}+3e^{4\lambda}).
\end{eqnarray}
In the expression for $G_{(4)}$, we have also used the relation $F^{12}=dA^{12}=\frac{1}{m}\textrm{vol}_{H^2}$. $\tilde{S}^2$ is a two-sphere inside the $S^4$ with coordinates $\hat{\mu}^{\hat{a}}$ and the volume form defined by 
\begin{equation}
\textrm{vol}_{\tilde{S}^2}=\frac{1}{2}\epsilon_{\hat{a}\hat{b}\hat{c}}\hat{\mu}^{\hat{a}}d\hat{\mu}^{\hat{b}}\wedge d\hat{\mu}^{\hat{c}}\, .
\end{equation}
Finally, the rotated vielbeins on $H^2$ are given by
\begin{equation}
\bar{e}^{'1}=\cos\theta \bar{e}^1+\sin\theta\bar{e}^2\qquad \textrm{and}\qquad \bar{e}^{'2}=\cos\theta \bar{e}^2-\sin\theta\bar{e}^1\, .
\end{equation}
\indent Near the $AdS_5$ geometry in the IR with $\tilde{r}\sim r\rightarrow -\infty$, the eleven-dimensional metric is given by
\begin{eqnarray}
ds^2_{11}&=&e^{\frac{2r}{L}}(ds^2_{AdS_2}+ds^2_{S^2})+dr^2+ds^2_{H^2}\nonumber \\
& &+\frac{1}{m^2}\left[d\xi^2+\cos^2\xi(d\theta-\omega_H)^2+\sin^2\xi d\hat{\mu}^{\hat{a}}d\hat{\mu}^{\hat{a}}\right]
\end{eqnarray}
which takes the form of a product between a locally $AdS_5$ geometry and an internal space given by $S^4$ fibration over $H^2$. The four-form field strength is given by
\begin{eqnarray}
G_{(4)}&=&\frac{3}{m^3}\sin^2\xi\cos\xi \textrm{vol}_{\tilde{S}^2}\wedge d\theta\wedge d\xi+\frac{1}{m^3}\textrm{vol}_{H^2}\wedge \textrm{vol}_{\tilde{S}^2}\nonumber \\
& &-\frac{1}{\sqrt{2}}e^{\frac{r}{L}}\cos\xi\left(\textrm{vol}_{S^2}\wedge dr\wedge \bar{e}^{'2}-\textrm{vol}_{AdS_2}\wedge dr\wedge \bar{e}^{'1}\right)\nonumber \\
& &+\frac{1}{\sqrt{2}}e^{\frac{r}{L}}\sin\xi d\xi\wedge \left(\textrm{vol}_{S^2}\wedge \bar{e}^{'2}-\textrm{vol}_{AdS_2}\wedge \bar{e}^{'1}\right)\nonumber \\
& &+\frac{1}{\sqrt{2}}e^{\frac{r}{L}}\cos\xi d\theta\wedge \left(\textrm{vol}_{S^2}\wedge \bar{e}^{'1}+\textrm{vol}_{AdS_2}\wedge \bar{e}^{'2}\right).\label{G4_2}
\end{eqnarray}
\indent At large $\tilde{r}$, the solution leads to the following asymptotic behavior
\begin{equation}
\phi_3\sim \ln \tilde{r},\qquad \Sigma\sim \tilde{r}^{-\frac{1}{3}},\qquad f\sim \frac{5}{3}\ln \tilde{r},\qquad b_1\sim \tilde{r}^2
\end{equation}
which gives
\begin{equation}
\phi\sim \frac{1}{3}\ln \tilde{r}\qquad \textrm{and}\qquad \lambda\sim\lambda_0\sim \textrm{constant}\, . 
\end{equation}
With all these, the eleven-dimensional metric becomes, see more detail in \cite{Janus_5D_ISO3},
\begin{eqnarray}
ds^2_{11}&\sim&\Delta_0^{\frac{1}{3}}\left[\tilde{r}^{2}\left(ds^2_{AdS_2}+ds^2_{S^2}+ds^2_{H_2}\right)+\frac{d\tilde{r}^2}{\tilde{r}^2}\right]+\frac{\Delta_0^{-\frac{2}{3}}}{m^2}\left[e^{6\lambda_0}\cos^2\xi (d\theta-\omega_H)^2\right.\nonumber \\
& &\left.+\left(e^{6\lambda_0}\sin^2\xi+ e^{-4\lambda_0}\cos^2\xi\right) d\xi^2+e^{-4\lambda_0}\sin^2\xi d\hat{\mu}^\alpha d\hat{\mu}^\alpha\right]
\end{eqnarray}
with $\Delta_0=e^{-6\lambda_0}\cos^2\xi+e^{4\lambda_0}\sin^2\xi$. Accordingly, the eleven-dimensional metric contains a locally $AdS_7$ geometry leading to the identification of the UV theory as the six-dimensional $N=(0,2)$ SCFT arising from M5-branes wrapped on $H^2$. In this limit, the four-form field strength takes the form
\begin{eqnarray}
G_{(4)}&=&-\frac{1}{m^3}\Delta_0^{-2}U_0\sin^2\xi\cos\xi \textrm{vol}_{\tilde{S}^2}\wedge d\theta\wedge d\xi+\frac{1}{m^3}\Delta_0^{-1}e^{4\lambda_0}\textrm{vol}_{H^2}\wedge \textrm{vol}_{\tilde{S}^2}\nonumber \\
& &-\frac{1}{\sqrt{2}}\tilde{r}e^{-6\lambda_0}\cos\xi\left(\textrm{vol}_{S^2}\wedge d\tilde{r}\wedge \bar{e}^{'2}-\textrm{vol}_{AdS_2}\wedge d\tilde{r}\wedge \bar{e}^{'1}\right)\nonumber \\
& &+\frac{1}{\sqrt{2}}\tilde{r}^2\sin\xi d\xi\wedge \left(\textrm{vol}_{S^2}\wedge \bar{e}^{'2}-\textrm{vol}_{AdS_2}\wedge \bar{e}^{'1}\right)\nonumber \\
& &+\frac{1}{\sqrt{2}}\tilde{r}^2\cos\xi d\theta\wedge \left(\textrm{vol}_{S^2}\wedge \bar{e}^{'1}+\textrm{vol}_{AdS_2}\wedge \bar{e}^{'2}\right).\label{G4_3}
\end{eqnarray}
with $U_0=2(e^{-6\lambda_0}\cos^2\xi+e^{4\lambda_0}\sin^2\xi)-\Delta_0(2e^{-6\lambda_0}+3e^{4\lambda_0})$.
\\
\indent From the four-form field strength, we find that for $b_1=0$, there are only M5-branes while non-vanishing $b_1$ leads to the presence of M2-branes in the configuration. Therefore, we expect the line defect solution would lead to a brane system with intersecting M2- and M5-branes. For the line defect in the form of charged domain walls from pure $N=4$ gauged supergravity studied in \cite{line_defects_5DN4_Nicolo}, the solution has been uplifted to type IIB theory in which it has been shown that the uplifted solution describes a near horizon limit of a brane configuration involving the intersection of F1-D1-D5-NS5 branes and the D3-KK system. It would be interesting to construct M2-M5-brane configuration whose near horizon limit is given by the geometry \eqref{uplifted_11Dmetric}. This might involve some modifications to the M2-M5-brane systems considered in \cite{Nicolo_deconstruct_defect,M2-M5_1,M2-M5_2} to describe surface defects in $N=(0,1)$ SCFT in six dimensions.   

\subsection{Solutions with $SO(2)\times SO(2)$ symmetry}
We end this section by considering another solution with $SO(2)\times SO(2)$ symmetry. In this case, we have $\phi_5\neq \phi_3$ with the BPS equations given by
\begin{eqnarray}
\phi'_3&=&\frac{1}{2}g\Sigma^{-1}e^{-2\phi_3-\phi_5}(e^{2\phi_5}-1),\\
\phi'_5&=&-\frac{1}{2}g\Sigma^{-1}e^{-2\phi_3-\phi_5}(1+2e^{2\phi_3}+e^{2\phi_5}),\\
\Sigma'&=&\frac{1}{6}\left[2\sqrt{2}g_1\Sigma^3+ge^{-2\phi_3-\phi_5}(e^{2\phi_5}+2e^{2\phi_3}-1)\right],\\
f'&=&\frac{1}{6}\Sigma^{-1}\left[ge^{-2\phi_3-\phi_5}(e^{2\phi_5}+2e^{2\phi_3}-1)-\sqrt{2}g_1\Sigma^3\right],\\
b_1&=&\frac{2\kappa}{g_1\Sigma}e^{f}\, .
\end{eqnarray}
The solutions for $\phi_3$, $\phi_5$, $\Sigma$ and $f$ have already been given in \cite{Janus_5D_ISO3}. Therefore, we will not repeat all the details here but simply give the result together with the solution for $b_1$. In terms of the scalar fields defined by   
\begin{equation}
\varphi_1=\phi_3+\phi_5\qquad \textrm{and}\qquad \varphi_2=\phi_3-\phi_5,\label{varphi12_def}
\end{equation}
the solution is given by
\begin{eqnarray}
f&=&\frac{1}{4}\varphi_2-\frac{1}{2}\ln \Sigma-\frac{3}{8}\ln(1-e^{2\varphi_2})+\frac{1}{4}\ln\left(\sin^{-1}e^{\varphi_2}-\frac{\pi}{2}\right),\nonumber \\
\Sigma^{-3}&=&\frac{g_1e^{-\varphi_2}\left(\sin^{-1}e^{\varphi_2}-\frac{\pi}{2}-e^{\varphi_2}\sqrt{1-e^{2\varphi_2}}\right)}{\sqrt{2}g(1-e^{2\varphi_2})^{\frac{1}{4}}\sqrt{\frac{\pi}{2}-\sin^{-1}e^{\varphi_2}}},\nonumber \\
e^{\varphi_1}&=&\frac{\frac{\pi}{2}-\sin^{-1}e^{\varphi_2}}{\sqrt{1-e^{2\varphi_2}}},\nonumber \\
2g(\tilde{r}-\tilde{r}_0)&=&\ln(1+e^{\frac{\varphi_2}{2}})-\ln(1-e^{\frac{\varphi_2}{2}})-2\tan^{-1}e^{\frac{\varphi_2}{2}},\nonumber \\
b_1&=&-\kappa 2^{\frac{3}{4}}\sqrt{\frac{\sin^{-1}e^{\varphi_2}-\frac{\pi}{2}-e^{\varphi_2}\sqrt{1-e^{2\varphi_2}}}{gg_1(e^{2\varphi_2}-1)}}
\end{eqnarray}
with the new radial coordinate $\tilde{r}$ defined by $\frac{d\tilde{r}}{dr}=\frac{e^{-\frac{\varphi_1}{2}}}{\Sigma}$ and a constant $\tilde{r}_0$. 
\\
\indent Near the $AdS_5$ vacuum, we find
\begin{eqnarray}
& &\Sigma\sim -\left(\frac{g}{\sqrt{2}g_1}\right)^{\frac{1}{3}}+e^{-\frac{2r}{L}},\qquad \varphi_2\sim e^{-\frac{4r}{L}},\nonumber \\
& &\varphi_1\sim \frac{1}{3}(2e^{\frac{2r}{L}}+e^{-\frac{4r}{L}}),\qquad b_1\sim \varphi_2^{-\frac{1}{4}}\sim e^{\frac{r}{L}}\, .
\end{eqnarray}
In terms of the original scalars $\phi_3$ and $\phi_5$, we have
\begin{equation}
2\phi_3+\phi_5\sim e^{\frac{2r}{L}}\qquad \textrm{and}\qquad \phi_3-\phi_5\sim e^{-\frac{4r}{L}}
\end{equation}
which indicates that there is a deformation by a source term of a dimension-6 operator dual to $2\phi_3+\phi_5$. Accordingly, the dual $N=2$ SCFT will appear in the IR. In \cite{Janus_5D_ISO3}, it has been shown that the UV limit corresponds to six-dimensional $N=(0,2)$ field theory on the world-volume of M5-branes wrapped on $H^2$. The corresponding eleven-dimensional metric in the UV limit has also been given in \cite{Janus_5D_ISO3}. However, in this case, the complete uplifted solution to eleven dimensions is much more complicated, so we refrain from giving it here. Due to some similarity between the two solutions, we again expect that the eleven-dimensional solution should describe a holographic line defect within the dual $N=2$ SCFT involving some intersection of M5- and M2-branes as in the previous $SO(2)\times SO(3)$ symmetric solution.  

\section{Line defects from $SO(2)\times SO(3)\times SO(3)$ gauged supergravity}\label{SO2_SO3_SO3}
As a final example, we consider line defect solutions from gauged supergravity with $SO(2)\times SO(3)\times SO(3)$ gauge group. This gauged supergravity admits two $N=4$ supersymmetric $AdS_5$ vacua with $SO(2)\times SO(3)\times SO(3)$ and $SO(2)\times SO(3)_{\textrm{diag}}$ symmetries and has been extensively studied in \cite{5D_N4_flow_Davide,5D_N4_flow}. Although this gauged supergravity currently has no known higher-dimensional origin, it turns out that the resulting solutions contain a line defect solution interpolating between the two aforementioned $AdS_5$ vacua. This type of solutions is similar to those studied in seven and six dimensions in \cite{7D_N2_DW_3_form,F4_defect} and has not appeared in the previous two sections. Accordingly, it could be useful to include these solutions in this paper.   
\\
\indent The $SO(2)\times SO(3)\times SO(3)$ gauged supergravity is obtained from coupling the supergravity multiplet to three vector multiplets. The embedding tensor takes the form
\begin{eqnarray}
\xi^{MN}&=&-g_1(\delta^M_1\delta^N_2-\delta^M_2\delta^N_1),\nonumber \\ 
f_{\tilde{m}+2,\tilde{n}+2,\tilde{p}+2}&=&h_1\epsilon_{\tilde{m}\tilde{n}\tilde{p}},\qquad \tilde{m},\tilde{n},\tilde{p}=1,2,3,\\
f_{\tilde{a}+5,\tilde{b}+5,\tilde{c}+5}&=&h_2\epsilon_{\tilde{a}\tilde{b}\tilde{c}},\qquad \tilde{a},\tilde{b},\tilde{c}=1,2,3
\end{eqnarray} 
with $g_1$, $h_1$ and $h_2$ being the gauge coupling constants. As in the previous sections, we consider $SO(2)_{\textrm{diag}}\sim [SO(2)\times SO(2)\times SO(2)]_{\textrm{diag}}\subset SO(2)\times SO(3)\times SO(3)$ symmetric scalar fields with the coset representative, see more detail in \cite{5D_N4_black_stringII}, given by
\begin{equation}
\mc{V}=e^{\phi_1(Y_{11}+Y_{22})}e^{\phi_2(Y_{12}-Y_{21})}e^{\phi_3(Y_{31}+Y_{42})}e^{\phi_4(Y_{32}-Y_{41})}e^{\phi_5Y_{53}}\, .
\end{equation}
As in the previous cases, we need to set $\phi_2=\phi_4=0$ in order to consistently truncate out all the vector fields.
\\
\indent With $\phi_2=\phi_4=0$, all non-vanishing dressed components of the embedding tensor are given by
\begin{eqnarray}
& &\hat{\xi}^{12}=-g_1\cosh^2\phi_1,\nonumber \\ 
& &\hat{\xi}^{34}=-g_1\sinh^2\phi_1\sinh^2\phi_3,\nonumber \\
& &\hat{\xi}^{14}=-\hat{\xi}^{23}=-\frac{1}{2}g_1\sinh2\phi_1\sinh\phi_3,\nonumber \\
& &\tilde{\xi}^{12}=-\tilde{\xi}^{21}=-\frac{1}{2}g_1\sinh2\phi_1\cosh\phi_3,\nonumber \\ 
& &\tilde{\xi}^{14}=-\tilde{\xi}^{23}=\frac{1}{2}g_1\cosh\phi_3\sinh2\phi_3,
\end{eqnarray}
\begin{eqnarray}
& &\hat{f}_{125}=-h_2\sinh^2\phi_1\sinh\phi_5,\nonumber \\
& & \hat{f}_{145}=-\hat{f}_{235}=-\frac{1}{2}h_2\sinh2\phi_1\sinh\phi_3\sinh\phi_5,\nonumber \\
& &\hat{f}_{345}=-h_1\cosh^2\phi_3\cosh\phi_5-h_2\cosh^2\phi_1\sinh^2\phi_3\sinh\phi_5,
\end{eqnarray}
\begin{eqnarray}
& &\tilde{f}^{125}=-\tilde{f}^{215}=\frac{1}{2}h_2\sinh2\phi_1\cosh\phi_3\sinh\phi_5,\nonumber \\
& &\tilde{f}^{145}=-\tilde{f}^{235}=\frac{1}{2}\sinh2\phi_3(h_1\cosh\phi_5+h_2\cosh^2\phi_1\sinh\phi_5),\nonumber \\
& &\tilde{f}^{312}=h_2\cosh\phi_5\sinh^2\phi_1,\nonumber \\
& &\tilde{f}^{314}=\tilde{f}^{323}=\frac{1}{2}h_2\sinh2\phi_1\cosh\phi_5\sinh\phi_3,\nonumber \\
& &\tilde{f}^{334}=h_1\cosh^2\phi_3\sinh\phi_5+h_2\cosh^2\phi_1\cosh\phi_5\sinh^2\phi_3
\end{eqnarray}
together with the vielbein on the $SO(5,3)/SO(5)\times SO(3)$ coset manifold
\begin{eqnarray}
& &{P^1}_m=-\frac{i}{2}(\cosh\phi_3\phi_1',0,\phi_3',0,0),\nonumber \\
& &{P^2}_m=-\frac{i}{2}(0,\cosh\phi_3\phi_1',0,\phi_3',0),\nonumber \\
& &{P^3}_m=-\frac{i}{2}(0,0,0,0,\phi_5').
\end{eqnarray}
As in the previous section, there are two two-form fields given by
\begin{equation}
B_1=b_1\textrm{vol}_{AdS_2}\qquad \textrm{and}\qquad B_2=\tilde{b}_2\textrm{vol}_{S^2}\, .
\end{equation}
\indent We then repeat the same analysis as in the previous two sections with a slightly different projector
\begin{equation}
\Gamma_5\tilde{\epsilon}=\tilde{\epsilon}\qquad \textrm{or}\qquad \Gamma_{14}\tilde{\epsilon}=-\Gamma_{23}\tilde{\epsilon}\, .
\end{equation}
This is due to some difference in the ordering of $SO(5,3)$ non-compact generators in the coset representative. We also note that this projector is not necessary for either $\phi_1=0$ or $\phi_3=0$ as in the previous sections. Similar to the case of $SO(2)\times ISO(3)$ gauge group, $\delta\lambda^3_i=0$ conditions lead to 
\begin{equation}
\phi_5'+\Sigma^{-1}(\tilde{f}^{312}+\tilde{f}^{334})=0\, .
\end{equation}
$\delta\lambda^1_i=0$ and $\delta\lambda^2_i=0$ conditions give rise to two algebraic constraints
\begin{equation}
\sinh\phi_1\tilde{b}_2=\sinh\phi_1b_1=0
\end{equation}
which imply that $\phi_1=0$ for non-vanishing two-form fields as in the previous section. In addition, $\delta\lambda^1_i$ and $\delta\lambda^2_i$ also give the BPS equation for non-vanishing $\phi_3$ provided that we choose $\theta=0$ or $\theta=\frac{\pi}{2}$. As in the previous sections, we will choose $\theta=0$ for convenience since both of these choices result in equivalent BPS equations. 
\\
\indent With $\phi_1=\theta=0$, $\delta\psi_{\alpha i}$ and $\delta\psi_{si}$ give $f=h$ as before. On the other hand, there is another algebraic constraint from $\delta\chi_i$ conditions of the form
\begin{equation}
b_1e^{-2f}=\tilde{b}_2e^{-2h}\, .
\end{equation}
After using $f=h$, we have $\tilde{b}_2=b_1$ as in all the previous cases. From $\delta\psi_{ri}$, we also have $\varphi'=0$ and $Y'=\frac{Y}{2}f'$. Choosing $\varphi=0$ and the same choice of signs $\kappa_i=\tilde{\kappa}_i=\kappa$, we find that all the remaining equations take a similar form as in the previous cases with different $\tilde{\xi}^{am}$, $\hat{\xi}^{mn}$, $\tilde{f}^{amn}$ and $\hat{f}_{mnp}$.
\\
\indent All the resulting BPS equations read
\begin{eqnarray}
\phi_3&=&-\frac{1}{2}\Sigma^{-1}\sinh2\phi_3(h_1\cosh\phi_5+h_2\sinh\phi_5),\\
\phi_5'&=&-\Sigma^{-1}(h_1\cosh^2\phi_3\sinh\phi_5+h_2\cosh\phi_5\sinh^2\phi_3),\\
\Sigma'&=&-\frac{\sqrt{2}}{3}g_1\Sigma^3+\frac{1}{3}(h_1\cosh^2\phi_3\cosh\phi_5+h_2\sinh^2\phi_3\sinh\phi_5),\\
f'&=&\frac{1}{3\sqrt{2}}g_1\Sigma^2+\frac{1}{3}\Sigma^{-1}(h_1\cosh^2\phi_3\cosh\phi_5+h_2\sinh^2\phi_3\sinh\phi_5),\\
b_1&=&\frac{2\kappa}{g_1\Sigma}e^{f},\\
\frac{Y'}{Y}&=&\frac{f'}{2}
\end{eqnarray}
It can be easily verified that compatibility between these BPS equations and the two-form field equation
\begin{equation}
b_1+\frac{1}{\sqrt{2}}g_1\Sigma^2b_1=0
\end{equation}
is automatic by the form of $\Sigma'$ and $f'$ equations.
\\
\indent Before giving line defect solutions, we briefly review the two $N=4$ supersymmetric $AdS_5$ vacua here for later convenience. The first $N=4$ vacuum has $SO(2)\times SO(3)\times SO(3)$ symmetry and is given by
\begin{equation}
\phi_3=\phi_5=0,\qquad \Sigma=\left(\frac{h_1}{\sqrt{2}g_1}\right)^{\frac{1}{3}},\qquad L_1=\frac{2}{h_1}\left(\frac{h_1}{\sqrt{2}g_1}\right)^{\frac{1}{3}}\, .
\end{equation}
We can also choose $h_1=\sqrt{2}g_1$ to have $\Sigma=1$ at the vacuum. The other $N=4$ vacuum preserves only $SO(2)\times SO(3)_{\textrm{diag}}$ symmetry and is given by
\begin{eqnarray}
& &\phi_3=\frac{1}{2}\ln\left[\frac{h_2-h_1}{h_2+h_1}\right],\qquad \Sigma^3=\frac{h_1h_2}{\sqrt{2}g_1\sqrt{h_2^2-h_1^2}},
\qquad
L_2=2^{\frac{5}{6}}\left(\frac{h_2^2-h_1^2}{g_1h_1^2h_2^2}\right)^{\frac{1}{3}}\nonumber \\
& &
\end{eqnarray}
with $\phi_5=\phi_3$.

\subsection{Solutions with $SO(2)\times SO(3)_{\textrm{diag}}$ symmetry}
We begin with a simple solution with $SO(2)\times SO(3)_{\textrm{diag}}$ symmetry by setting $\phi_5=\phi_3$. The BPS equations become
\begin{eqnarray}
\phi_3'&=&-\frac{1}{2}\Sigma^{-1}(h_1\cosh\phi_3+h_2\sinh\phi_3)\sinh2\phi_3,\nonumber \\
\Sigma'&=&-\frac{\sqrt{2}}{3}g_1\Sigma^3+\frac{1}{3}(h_1\cosh^3\phi_3+h_2\sinh^3\phi_3),\nonumber \\
f'&=&\frac{\sqrt{2}}{6}g_1\Sigma^2+\frac{1}{3}\Sigma^{-1}(h_1\cosh^3\phi_3+h_2\sinh^3\phi_3),\nonumber \\
b_1&=&\frac{2\kappa}{g_1\Sigma}e^{f}\, .
\end{eqnarray}
The solutions for $\phi_3$, $\Sigma$ and $f$ have been analytically given in \cite{5D_N4_flow_Davide,5D_N4_flow}. Using this result to determine the two-form field $b_1$, we end up with the line defect solution
\begin{eqnarray}
\Sigma^3&=&\frac{h_1h_2}{\sqrt{2}g_1(h_1\sinh\phi_3+h_2\cosh\phi_3)},\nonumber \\
f&=&\frac{1}{2}\ln\left[\frac{h_1\cosh\phi_3+h_2\sinh\phi_3}{\sinh2\phi_3}\right]+\frac{1}{6}\ln(h_1\sinh\phi_3+h_2\cosh\phi_3),\nonumber \\
b_1&=&\kappa\left(\frac{4}{g_1^2h_1h_2}\right)^{\frac{1}{3}}\sqrt{h_1^2+h_2^2+h_1h_2(\tanh\phi_3+\coth\phi_3)},\nonumber \\
h_1h_2\tilde{r}&=&2h_1\tan^{-1}\tanh\frac{\phi_3}{2}-h_2\ln\tanh\frac{\phi_3}{2}\nonumber \\
& &-\sqrt{h_2^2-h_1^2}\tanh^{-1}\left[\frac{h_2+h_1\tanh\frac{\phi_3}{2}}{\sqrt{h_2^2-h_1^2}}\right]
\end{eqnarray} 
with $\tilde{r}$ defined by $\frac{d\tilde{r}}{dr}=\Sigma^{-1}$. In the solution, we have chosen an integration constant in $\Sigma$ such that the solution interpolates between the two $N=4$ supersymmetric $AdS_5$ vacua, see more detail in \cite{5D_N4_flow_Davide,5D_N4_flow}. In addition, we have also neglected additive integration constants in $f$ and $\tilde{r}$. 
\\
\indent Near the $SO(2)\times SO(3)\times SO(3)$ symmetric $AdS_5$ vacuum as $r\rightarrow \infty$, we find
\begin{equation}
\phi_3\sim e^{-\frac{2r}{L_1}},\qquad \Sigma\sim \left(\frac{h_1}{\sqrt{2}g_1}\right)^{\frac{1}{3}}\left(1-\frac{h_1}{3h_2}e^{-\frac{2r}{L_1}}\right),\qquad b_1\sim e^{\frac{r}{L_1}}
\end{equation}
which indicates that both $\phi_3$ and $\Sigma$ correspond to vacuum expectation values of two dual operators of dimension $\Delta=2$. In addition, as in all the previous cases, there are deformations by operators of dimension $\Delta=3$ dual to the two-form fields $B_1$ and $B_2$.
\\
\indent As $r\rightarrow -\infty$, the solution is asymptotic to the $SO(2)\times SO(3)_{\textrm{diag}}$ symmetric $AdS_5$ vacuum with
\begin{equation}
\phi_3\sim e^{\frac{2r}{L_2}},\qquad \Sigma-\left(\frac{h_2}{\sqrt{h_2^2-h_1^2}}\right)^{\frac{1}{3}}\sim e^{-\frac{2r}{L_2}},\qquad b_1\sim e^{\frac{r}{L_2}}\, .
\end{equation}
In this limit, $\phi_3$ is now dual to an operator of dimension $\Delta=6$ while $\Sigma$ and $b_1$ are still dual to operators of dimensions $\Delta=2$ and $\Delta=3$, respectively.

\subsection{Solutions with $SO(2)\times SO(2)\times SO(2)$ symmetry}
Unlike the $SO(2)\times ISO(3)$ gauged supergravity studied in the previous section, we are not able to analytically solve the BPS equations with $\phi_5\neq\phi_3$ in full generality. Although a numerical analysis would still lead to some interesting solutions, it could be more useful to have analytic solutions in certain specific cases. To give an example of these solutions, we consider solutions with $SO(2)\times SO(2)\times SO(2)$ symmetry obtained by setting $\phi_3=0$. In this case, we find the solution of the form
\begin{eqnarray}
\Sigma^3&=&\frac{h_1}{\sqrt{2}g_1\cosh\phi_5+h_1C_0\sinh\phi_5},\nonumber \\
f&=&-\frac{1}{2}\ln\sinh\phi_5+\frac{1}{6}\ln(\sqrt{2}g_1\cosh\phi_5+h_1C_0\sinh\phi_5),\nonumber \\
b_1&=&\frac{2\kappa}{g_1h_1^{\frac{1}{3}}}\sqrt{\frac{\sqrt{2}g_1\cosh\phi_5+h_1C_0\sinh\phi_5}{\sinh\phi_5}},\nonumber \\
h_1\tilde{r}&=&\ln\cosh\frac{\phi_5}{2}-\ln\sinh\frac{\phi_5}{2}
\end{eqnarray} 
with $\tilde{r}$ defined by $\frac{d\tilde{r}}{dr}=\Sigma^{-1}$ and $C_0$ being an integration constant. This solution is analogous to all the solutions given in the previous sections in a sense that the solution interpolates between the $N=4$ supersymmetric $AdS_5$ vacuum with $SO(2)\times SO(3)\times SO(3)$ symmetry at the origin of the scalar manifold and a singular geometry. 
\\
\indent Near the $AdS_5$ vacuum, we find a similar asymptotic behavior as in the previous solutions
\begin{equation}
\phi_5\sim e^{-\frac{2r}{L_1}},\qquad \Sigma\sim \left(\frac{h_1}{\sqrt{2}g_1}\right)^{\frac{1}{3}}\left(1-\frac{h_1}{3\sqrt{2}g_1}C_0e^{-\frac{2r}{L_1}}\right),\qquad b_1\sim e^{\frac{r}{L_1}}\, .
\end{equation}
We see again that the constant $C_0$ controls the vacuum expectation value of the dimension-2 operator dual to $\Sigma$. In the previous solution, this constant has been tuned to make the solution approach the second $N=4$ supersymmetric $AdS_5$ vacuum. 
\\
\indent At $\tilde{r}=0$, the solution is singular with
\begin{equation}
\phi_5\sim -\frac{1}{2}\ln\tilde{r}\, .
\end{equation}
The behaviours of all the other field depend on the values of $C_0$. For $C_0\neq -\frac{\sqrt{2}g_1}{h_1}$, we find
\begin{equation}
\Sigma\sim \tilde{r}^{\frac{1}{6}},\qquad f\sim \frac{1}{6}\ln\tilde{r},\qquad b_1\sim \textrm{constant}
\end{equation}
while, for $C_0= -\frac{\sqrt{2}g_1}{h_1}$, the solution becomes
\begin{equation}
\Sigma\sim \tilde{r}^{-\frac{1}{6}},\qquad f\sim \frac{1}{3}\ln\tilde{r},\qquad b_1\sim \sqrt{\tilde{r}}\, .
\end{equation}
We can use the scalar potential given in \cite{5D_N4_flow} to determine whether these singularities are physically acceptable by the criterion of \cite{Gubser_Sing} or not. It turns out that both of these singularities lead to $V\rightarrow -\infty$, so both singularities are physical. However, as previously mentioned, the higher-dimensional origin of this gauged supergravity is unknown. Accordingly, we do not know whether these singularities lead to any sensible solutions in string/M-theory.
\section{Conclusions and discussions}\label{conclusion}
We have studied a number of holographic solutions describing conformal line defects within $N=2$ SCFTs from five-dimensional $N=4$ gauged supergravity with $SO(2)_D\times SO(3)$, $SO(2)\times ISO(3)$ and $SO(2)\times SO(3)\times SO(3)$ gauge groups. 
All of the solutions take the form of charged domain walls with $AdS_2\times S^2$ slices. In these solutions, the warp factors for both $AdS_2$ and $S^2$ are equal, and the phase $\theta$ of the Killing spinors must be $0$ or $\frac{\pi}{2}$. This is in constrast to the results from pure $N=4$ gauged supergravity in which solutions with a non-trivial phase $\theta(r)$ of the Killing spinors and different warp factors of $AdS_2$ and $S^2$ exist \cite{line_defects_5DN4_Nicolo}. As shown in the previous three sections, the BPS equations of the scalar fields from vector multiplets impose a strong restriction on the phase $\theta(r)$ of the Killing spinors in such a way that $\theta$ must be $0$ or $\frac{\pi}{2}$. The latter in turn lead to the same warp factors for both $AdS_2$ and $S^2$. This is similar to solutions from six- and seven-dimensional gauged supergravities studied in \cite{7D_N2_DW_3_form,F4_defect,ISO3_defect}. All of these results seem to suggest that holographic solutions describing conformal defects from half-maximal gauged supergravities coupled to vector multiplets can only take the form of charged domain walls with equal warp factors for both the anti-de Sitter and sphere parts of the metric. This could presumably stem from the nanner in which the two- and three-form tensor fields couple to scalars from vector multiplets.      
\\
\indent One possibility to resolve this issue could be to turn on the $SO(2)_{\textrm{diag}}$ gauge field in addition to the two-form fields considered here. We can take the gauge field strength tensors of the form
\begin{equation}
F^{\mc{M}}=a_{\mc{M}}\textrm{vol}_{AdS_2}+\tilde{a}_{\mc{M}}\textrm{vol}_{S^2}\, .
\end{equation}
For $SO(2)_D\times SO(3)$ and $SO(2)\times ISO(3)$ gauge groups, the $SO(2)_{\textrm{diag}}$ gauge field corresponds to a linear combination of $A^0$ and $A^3$. In $SO(2)\times SO(3)\times SO(3)$ gauge group, this gauge field would be identified with a certain combination of $A^0$ and $A^5$. The non-vanishing gauge field leads to an extra contribution in $\delta \lambda^a_i$ of the form
\begin{equation}
-\frac{1}{4}\Sigma {\mc{V}_M}^aF^M_{\mu\nu}\gamma^{\mu\nu}\epsilon_i
\end{equation} 
for $M=3$ or $M=5$. We also note that $A^0$ does not appear in $\delta\lambda^a_i$. In addition, for non-vanishing gauge fields, it might be necessary to retain all the $SO(2)_{\textrm{diag}}$ singlet scalars involving non-vanishing $\phi_{2}$ and $\phi_4$ and relax the constraint $d_{MNP}\xi^{MQ}\xi^{NR}B_Q\wedge B_R=0$. Consequently, we could not expect to analytically solve the resulting BPS equations. In $SO(2)_D\times SO(3)$ gauge group, we have ${\mc{V}_3}^{6,7}=0$ even though $\phi_2\neq 0$ and $\phi_4\neq 0$, so the gauge field $A^3$ does not contribute to the BPS equations of the scalars from vector multiplets. This again implies that we need to set $\theta=0$ or $\theta=\frac{\pi}{2}$ and eventually leads to solutions in the form of charged dimain walls with equal warp factors for $AdS_2$ and $S^2$. On the other hand, for the remaining two gauge groups, we have respectively ${\mc{V}_3}^6=\sinh\phi_5$ and ${\mc{V}_5}^8=\sinh\phi_5$ with all the other components vanishing. Therefore, in these gauge groups, the BPS equations for scalar fields from vector multiplets could have additional contributions from the $SO(2)_{\textrm{diag}}$ gauge field. It would be interesting to investigate this issue in more detail. However, various equations are also modified and become much more complicated.  
\\
\indent In $SO(2)_D\times SO(3)$ gauge group, we have not found any solutions that are asymptotic to the $N=2$ supersymmetric $AdS_5$ vacuum due to the inconsistency between the BPS equations and the two-form field equations requiring the gauge coupling constant $g_2$ to vanish. Therefore, there is no line defect solution interpolating between $N=4$ and $N=2$ $AdS_5$ vacua. This in turn implies that all the solutions found in this case are solutions of $N=4$ gauged supergravity coupled to vector multiplets with $SO(2)\times SO(3)$ gauge group. The result can then be considered as an extension to the charged domain wall solution studied recently in \cite{line_defects_5DN4_Nicolo} from pure $N=4$ gauged supergravity with $SO(2)\times SO(3)$ gauge group.   
\\
\indent For $SO(2)\times ISO(3)$ gauge group, there is only one $N=4$ supersymmetric $AdS_5$ vacuum arising from the near horizon geometry of M5-branes wrapped on $H^2$. In this case, some of the scalars from vector multiplets appearing in the solutions are dual to irrelevant operators, so the presence of these deformations implies that the dual $N=2$ SCFT appears in the IR. The UV theories have been argued recently in \cite{Janus_5D_ISO3} to be $N=(0,2)$ six-dimensional field theories arising from M5-branes wrapped on $H^2$. The solutions given in this paper are accordingly interpreted as such RG flows in the presence of line defects generated by turning on the deformations involving dimension-3 operators dual to the two-form fields. The uplifted solution with $SO(2)\times SO(3)$ symmetry suggests that the defects should arise from a brane configuration involving some intersection of M2- and M5-branes.     
\\
\indent In $SO(2)\times SO(3)\times SO(3)$ gauge group, there exists a solution describing holographic line defects that interpolates between two $N=4$ supersymmetric $AdS_5$ vacua with $SO(2)\times SO(3)\times SO(3)$ and $SO(2)\times SO(3)_{\textrm{diag}}$ symmetries. This extends the RG flow solution given in \cite{5D_N4_flow_Davide} and \cite{5D_N4_flow} by turning on non-vanishing two-form fields leading to line defects within the dual $N=2$ SCFTs. This type of solutions have also been given in half-maximal gauged supergravities in six and seven dimensions in \cite{7D_N2_DW_3_form,F4_defect}. In addition, we have given a solution with $SO(2)\times SO(2)\times SO(2)$ symmetry that is only asymptotic to the $N=4$ $AdS_5$ vacuum with $SO(2)\times SO(3)\times SO(3)$ symmetry. Although this gauged supergravity currently has no known higher-dimensional origin, these solutions could be of interest in the study of holographic line defects within $N=2$ SCFTs.   
\\
\indent It would be interesting to use holographic renormalization to compute on-shell action of the five-dimensional gauged supergravity considered here and determine correlation functions in the dual $N=2$ SCFTs in the presence of line defects similar to a recent result in \cite{line_surface_Nicolo_new}. Embedding the solutions from $SO(2)_D\times SO(3)$ and $SO(2)\times SO(3)\times SO(3)$ gauge groups in string/M-theory is also worth considering. For $SO(2)_D\times SO(3)$ gauge group, this obviously involves solving algebraic constraints presented in \cite{Malek_AdS5_N4_embed} and determine explicit forms of the corresponding truncation ansatze for various fields. On the other hand, embedding the $SO(2)\times SO(3)\times SO(3)$ gauged supergravity in higher dimensions would be more challenging and highly desirable since this could lead to a higher-dimensional interpretation of the line defect solution interpolating between two different $N=2$ SCFTs. Furthermore, the resulting truncation ansatz can also be used to uplift a large number of interesting holographic solutions studied in \cite{5D_N4_flow_Davide,5D_flowII,5D_N4_flow,5D_N4_flow1,5D_N4_black_stringII,5D_Janus_new} to higher dimensions. We hope to come back to these issues in future work.
\vspace{0.5cm}\\
{\large{\textbf{Acknowledgement}}} \\
The author would like to thank Nicolo Petri for helpful correspondence.
\appendix
\section{Detailed analysis of the BPS equations}
In this appendix, we give more details on the analysis of the BPS equations for conformal line defect solutions to $N=4$ gauged supergravity in five dimensions. The relevant procedure is essentially the same in all gauge groups since the results can be written exclusively in terms of the dressed components of the embedding tensor. Accordingly, we will present the analysis only in the case of $SO(2)_D\times SO(3)$ gauge group.
\\
\indent Using the ansatz for the Killing spinors in \eqref{Killing_spin} which for convenience we also repeat here
\begin{equation} 
\epsilon_i=Y\left[\cos\theta {\delta_i}^j+\sin\theta \gamma_{01}{(\Gamma_1)_i}^j\right]\tilde{\epsilon}_j,
\end{equation}
we find that $\delta\lambda^1_i=0$ condition leads to
\begin{eqnarray}
0&=&-\frac{1}{2}(\cosh\phi_3\phi_1'\Gamma_1+\phi_3'\Gamma_4)(\cos\theta+\sin\theta \gamma_{01}\Gamma_1)i\gamma^{\hat{r}}\tilde{\epsilon}\nonumber \\
& &-\frac{1}{2}\left\{\frac{1}{\sqrt{2}}\Sigma^2(\tilde{\xi}^{12}\Gamma_2+\tilde{\xi}^{15}\Gamma_{5})+\Sigma^{-1}\tilde{f}^{135}\Gamma_{35}\right\}(\cos\theta+\sin\theta\gamma_{01}\Gamma_1)\tilde{\epsilon}\nonumber \\
& &-\frac{1}{4}\Sigma e^{-2h}\coth\phi_3K_4\gamma_{01}(\cos\theta+\sin\theta\gamma_{01}\Gamma_1)i\gamma_{\hat{r}}\tilde{\epsilon}
\end{eqnarray}
in which we have omitted $SO(5)$ spinor indices $i,j$ for brevity. We also recall that $K_4$ is defined in \eqref{K4_K5_def}. 
\\
\indent We now use the projector
\begin{equation}
\gamma_{\hat{r}}\tilde{\epsilon}=i\Gamma_{12}\tilde{\epsilon},\label{gamma_rprodap}
\end{equation}  
and the above equation becomes
\begin{eqnarray}
0&=&\frac{1}{2}(\cosh\phi_3\phi_1'\Gamma_1+\phi_3'\Gamma_4)(\cos\theta\Gamma_{12}+\sin\theta \gamma_{01}\Gamma_2)\tilde{\epsilon}\nonumber \\
& &-\frac{1}{2}\left\{\frac{1}{\sqrt{2}}\Sigma^2(\tilde{\xi}^{12}\Gamma_2+\tilde{\xi}^{15}\Gamma_5)+\Sigma^{-1}\tilde{f}^{135}\Gamma_{35}\right\}(\cos\theta+\sin\theta\gamma_{01}\Gamma_1)\tilde{\epsilon}\nonumber \\
& &+\frac{1}{4}\Sigma e^{-2h}\coth\phi_3K_4(\cos\theta\gamma_{01}\Gamma_{12}+\sin\theta\Gamma_2)\tilde{\epsilon}\, .
\end{eqnarray}   
The coefficient of $\Gamma_2$ gives 
\begin{equation}
0=\frac{1}{2}\left(\cosh\phi_3\phi_1'-\frac{1}{\sqrt{2}}\Sigma^2\tilde{\xi}^{12}\right)\cos\theta+\frac{1}{4}\Sigma e^{-2h}\coth\phi_3 K_4\sin\theta\, .
\end{equation}        
Similarly, the coefficient of $\gamma_{01}\Gamma_{12}$ leads to
\begin{equation}
0=\frac{1}{2}\left(\cosh\phi_3\phi_1'+\frac{1}{\sqrt{2}}\Sigma^2\tilde{\xi}^{12}\right)\sin\theta+\frac{1}{4}\Sigma e^{-2h}\coth\phi_3 K_4\cos\theta\, .     
\end{equation}
Using $\Gamma_{12345}=\mathbb{I}_4$ which implies $\Gamma_{124}\tilde{\epsilon}=\Gamma_{35}\tilde{\epsilon}$ and a projector involvng $\Gamma_3$ given in \eqref{Gamma3_pro_app}, we find the following equations from the coefficients of $\Gamma_{124}$ and $\gamma_{01}\Gamma_{24}$
\begin{eqnarray}
0&=&\frac{1}{2}\left(\phi_3'-\Sigma^{-1}\tilde{f}^{135}-\frac{1}{\sqrt{2}}\Sigma^2\tilde{\xi}^{15}\right)\cos\theta,\nonumber \\
0&=&\frac{1}{2}\left(\phi_3'+\Sigma^{-1}\tilde{f}^{135}-\frac{1}{\sqrt{2}}\Sigma^2\tilde{\xi}^{15}\right)\sin\theta\, .
\end{eqnarray}     
\indent We then repeat the same procedure for $\delta\lambda^2_i=0$ which reads
\begin{eqnarray}
0&=&\frac{1}{2}(\cosh\phi_3\phi_1'\Gamma_2+\phi_3'\Gamma_5)(\cos\theta\Gamma_{12}+\sin\theta\gamma_{01}\Gamma_2)\tilde{\epsilon}\nonumber \\
& &+\frac{1}{2}\left(\Sigma\tilde{f}^{135}\Gamma_{34}+\frac{1}{\sqrt{2}}\Sigma^2\tilde{\xi}^{15}\Gamma_4+\frac{1}{\sqrt{2}}\Sigma^2\tilde{\xi}^{12}\Gamma_1\right)(\cos\theta+\sin\theta\gamma_{01}\Gamma_1)\tilde{\epsilon}\nonumber \\
& &+\frac{1}{4}\Sigma e^{-2f}\coth\phi_3K_5(\cos\theta\gamma_{01}+\sin\theta\Gamma_1)\tilde{\epsilon}               
\end{eqnarray}
with $K_5$ defined in \eqref{K4_K5_def}. Setting the coefficients of $\Gamma_1$, $\gamma_{01}$, $\Gamma_{125}$ and $\gamma_{01}\Gamma_{25}$ to zero, we obtain
\begin{eqnarray}
0&=&\frac{1}{2}\left(-\cosh\phi_3\phi_1'+\frac{1}{\sqrt{2}}\Sigma^2\tilde{\xi}^{12}\right)\cos\theta+\frac{1}{4}\Sigma e^{-2f}\coth\phi_3K_5\sin\theta,\nonumber \\
0&=&\frac{1}{2}\left(\cosh\phi_3\phi_1'+\frac{1}{\sqrt{2}}\Sigma^2\tilde{\xi}^{12}\right)\sin\theta+\frac{1}{4}\Sigma e^{-2f}\coth\phi_3K_5\cos\theta,\nonumber \\
0&=&\frac{1}{2}\left(-\phi_3'+\Sigma\tilde{f}^{135}+\frac{1}{\sqrt{2}}\Sigma^2\tilde{\xi}^{15}\right)\cos\theta,\nonumber \\
0&=&\frac{1}{2}\left(\phi_3'+\Sigma\tilde{f}^{135}-\frac{1}{\sqrt{2}}\Sigma^2\tilde{\xi}^{15}\right)\cos\theta\, .
\end{eqnarray}
\indent We then move to $\delta\chi_i$ variation which gives
\begin{eqnarray}
0&=&-\frac{\sqrt{3}}{2}i\frac{\Sigma'}{\Sigma}\gamma^{\hat{r}}\epsilon-\frac{1}{\sqrt{6}}\Sigma^2(\hat{\xi}^{12}\Gamma_{12}+\hat{\xi}^{15}\Gamma_{15}+\hat{\xi}^{24}\Gamma_{24}+\hat{\xi}^{45}\Gamma_{45})\epsilon
\nonumber\\
& &-\frac{1}{2\sqrt{3}\Sigma^{-1}}\hat{f}_{345}\Gamma^{345}\epsilon+\frac{1}{4\sqrt{3}}\Sigma e^{-2h}(K_1\Gamma_1+K_4\Gamma_4)\gamma^{34}\epsilon\nonumber \\
& &+\frac{1}{4\sqrt{3}}\Sigma e^{-2f}(K_2\Gamma_2+K_5\Gamma_5)\gamma^{01}\epsilon\, .
\end{eqnarray}
We also recall from \eqref{hat_xi_def1} that $\hat{\xi}^{24}=-\hat{\xi}^{15}$, so to avoid too many algebraic constraints, we will impose an additional projector
\begin{equation}
\Gamma_3\tilde{\epsilon}=-\tilde{\epsilon}\label{Gamma3_pro_app}
\end{equation}
which implies $\Gamma_{12}\tilde{\epsilon}=\Gamma_{45}\tilde{\epsilon}$. Furthermore, from the explicit expressions of $\hat{\xi}^{24}$ and $\hat{\xi}^{15}$, we see that both $\hat{\xi}^{24}$ and $\hat{\xi}^{15}$ vanish when either $\phi_1=0$ or $\phi_3=0$. Therefore, a general solution with non-vanishing $\phi_1$ and $\phi_3$ would preserve $\frac{1}{4}$ of the original supersymmetry while solutions with only $\phi_1\neq 0$ or $\phi_3\neq 0$ are half-supersymmetric.
\\
\indent With this extra projector and the usual $\gamma_{\hat{r}}$ projector in \eqref{gamma_rprodap}, we find
\begin{eqnarray}
0&=&\frac{\sqrt{3}}{2}\frac{\Sigma'}{\Sigma}(\cos\theta\Gamma_{12}+\sin\theta\gamma_{01}\Gamma_2)\tilde{\epsilon}\nonumber \\
& &+\frac{1}{2\sqrt{3}}\left\{\Sigma^{-1}\hat{f}_{345}-\sqrt{2}\Sigma^2(\hat{\xi}^{12}+\hat{\xi}^{45})\right\}(\cos\theta\Gamma_{12}-\sin\theta \gamma_{01}\Gamma_2)\tilde{\epsilon}\nonumber \\
& &-\frac{1}{4\sqrt{3}}\Sigma e^{-2h}\left\{(K_1\Gamma_{12}-K_4\Gamma_{15})\sin\theta+(K_1\Gamma_2+K_4\Gamma_5)\cos\theta\gamma_{01}\right\}\tilde{\epsilon}\nonumber \\
& &-\frac{1}{4\sqrt{3}}\Sigma e^{-2f}\left\{(K_2\Gamma_2+K_5\Gamma_5)\cos\theta\gamma_{01}-(K_2\Gamma_{12}+K_5\Gamma_{15})\sin\theta\right\}\tilde{\epsilon}\, .\qquad
\end{eqnarray}
Setting the coefficients of $\Gamma_{12}$, $\Gamma_2\gamma_{01}$, $\Gamma_{01}\Gamma_5$ and $\Gamma_{15}$ to zero gives
\begin{eqnarray}
0&=&\left\{\frac{\sqrt{3}}{2}\frac{\Sigma'}{\Sigma}+\frac{1}{2\sqrt{3}}\Sigma^{-1}\hat{f}_{345}-\frac{1}{\sqrt{6}}\Sigma^2(\hat{\xi}^{12}+\hat{\xi}^{45})\right\}\cos\theta\nonumber \\
& &+\frac{1}{4\sqrt{3}}\Sigma(K_2e^{-2f}-K_1e^{-2h})\sin\theta,\nonumber \\
0&=&\left\{\frac{\sqrt{3}}{2}\frac{\Sigma'}{\Sigma}-\frac{1}{2\sqrt{3}}\Sigma^{-1}\hat{f}_{345}+\frac{1}{\sqrt{6}}\Sigma^2(\hat{\xi}^{12}+\hat{\xi}^{45})\right\}\sin\theta\nonumber \\
& &-\frac{1}{4\sqrt{3}}\Sigma(K_2e^{-2f}+K_1e^{-2h})\cos\theta,\nonumber \\
0&=&(e^{-2h}K_4+e^{-2f}K_5)\cos\theta=(e^{-2h}K_4+e^{-2f}K_5)\sin\theta\, .
\end{eqnarray}
\indent To perform the analysis of the gravitino variations, it is useful to recall the metric ansatz
\begin{equation}
ds^2=e^{2f}ds^2_{AdS_2}+dr^2+e^{2h}ds^2_{S^2}\, .
\end{equation}
Using the vielbein of the form
\begin{equation}
e^{\hat{\alpha}}=e^f\hat{e}^{\hat{\alpha}},\qquad e^{\hat{r}}=dr,\qquad e^{\hat{s}}=e^h\tilde{e}^{\hat{s}}
\end{equation}
with $\hat{e}^{\hat{\alpha}}$ and $\tilde{e}^{\hat{s}}$ being respectively the vielbeins on $AdS_2$ and $S^2$, we find the following non-vanishing components of the spin connection
\begin{eqnarray}
{\omega^{\hat{\alpha}}}_{\hat{r}}=f'e^{\hat{\alpha}},\qquad {\omega^{\hat{\alpha}}}_{\hat{\beta}}={\hat{\omega}^{\hat{\alpha}}}_{\phantom{\hat{\alpha}}\hat{\beta}},\qquad{\omega^{\hat{s}}}_{\hat{r}}=h'e^{\hat{s}},\qquad {\omega^{\hat{s}}}_{\hat{u}}={\tilde{\omega}^{\hat{s}}}_{\phantom{\hat{s}}\hat{u}}\, .
\end{eqnarray}
All of these lead to the covariant derivatives of the Killing spinors along $AdS_2$ and $S^2$ of the form
\begin{equation}
D_{\hat{\alpha}}=e^{-f}\hat{\nabla}_{\hat{\alpha}}\epsilon+\frac{1}{2}f'\gamma_{\hat{\alpha}\hat{r}}\epsilon\qquad \textrm{and}\qquad 
D_{\hat{s}}=e^{-h}\tilde{\nabla}_{\hat{s}}\epsilon+\frac{1}{2}h'\gamma_{\hat{s}\hat{r}}\epsilon
\end{equation}
with $\hat{\nabla}_\alpha$ and $\tilde{\nabla}_s$ being the covariant derivatives on $AdS_2$ and $S^2$, respectively.
\\
\indent Using the Killing spinor equation on $AdS_2$ given in \eqref{AdS2_S2_Killing}, we find the following equation from $\delta\psi_{\alpha i}$ 
\begin{eqnarray}
0&=&\frac{f'}{2}(\cos\theta\Gamma_{12}+\sin\theta\gamma_{01}\Gamma_2)(\eta\otimes \chi)+\frac{\kappa}{2}e^{-f}e^{-2i\varphi\gamma_{\hat{r}}}(\cos\theta\Gamma_{12}\gamma_{01}-\sin\theta\Gamma_2)(\eta\otimes \chi)\nonumber \\
& &+\frac{1}{6}\left\{\frac{1}{\sqrt{2}}\Sigma^2(\hat{\xi}^{12}+\hat{\xi}^{45})+\Sigma^{-1}\hat{f}_{345}\right\}(\cos\theta\Gamma_{12}-\sin\theta\gamma_{01}\Gamma_2)(\eta\otimes \chi)\nonumber \\
& &+\frac{1}{6}\Sigma e^{-2f}(K_2\Gamma_2+K_5\Gamma_5)(\cos\theta\gamma_{01}+\sin\theta\Gamma_1)(\eta\otimes \chi)\nonumber \\
& &-\frac{1}{12}\Sigma e^{-2h}(K_1\Gamma_1+K_4\Gamma_4)(\cos\theta\gamma_{01}\Gamma_{12}+\sin\theta\Gamma_2)(\eta\otimes \chi).
\end{eqnarray}
Upon setting the coefficients of $\Gamma_{12}$, $\gamma_{01}\Gamma_2$, $\Gamma_{15}$ and $\Gamma_{5}\gamma_{01}$ to zero, we find
\begin{eqnarray}
0&=&\frac{1}{2}\left\{f'+\frac{1}{3\sqrt{2}}\Sigma^2(\hat{\xi}^{12}+\hat{\xi}^{45})+\frac{1}{3}\Sigma^{-1}\hat{f}_{345}\right\}\cos\theta(\eta\otimes \chi)\nonumber \\
& &-\frac{1}{2}\left\{\kappa e^{-f}e^{-2i\varphi\gamma_{\hat{r}}}\Gamma_1+\frac{1}{6}\Sigma(K_1e^{-2h}+2K_2e^{-2f})\right\}\sin\theta(\eta\otimes \chi),\nonumber \\
0&=&-\frac{1}{2}\left\{-f'+\frac{1}{3\sqrt{2}}\Sigma^2(\hat{\xi}^{12}+\hat{\xi}^{45})-\frac{1}{3}\Sigma^{-1}\hat{f}_{345}\right\}\sin\theta(\eta\otimes \chi)\nonumber \\
& &+\frac{1}{2}\left\{\kappa e^{-f}e^{-2i\varphi\gamma_{\hat{r}}}\Gamma_1-\frac{1}{6}\Sigma(K_1e^{-2h}-2K_2e^{-2f})\right\}\cos\theta(\eta\otimes \chi),\nonumber \\
0&=&(e^{-2h}K_4-2e^{-2f}K_5)\sin\theta=(e^{-2h}K_4-2e^{-2f}K_5)\cos\theta\, .
\end{eqnarray}
Similarly, by repeating the same analysis for $\delta\psi_{s i}$ variations, we obtain
\begin{eqnarray}
0&=&\frac{h'}{2}(\cos\theta\Gamma_{12}+\sin\theta\gamma_{01}\Gamma_2)(\eta\otimes \chi)-\frac{\tilde{\kappa}}{2}e^{-h}e^{-2i\varphi\gamma_{\hat{r}}}(\cos\theta\gamma_{01}\Gamma_{12}+\sin\theta\Gamma_2)(\eta\otimes \chi)\nonumber \\
& &+\frac{1}{6}\left\{\frac{1}{\sqrt{2}}\Sigma^2(\hat{\xi}^{12}\Gamma_{12}+\hat{\xi}^{45}\Gamma_{45})-\Sigma^{-1}\hat{f}_{345}\Gamma^{345}\right\}(\cos\theta+\sin\theta\gamma_{01}\Gamma_1)(\eta\otimes \chi)\nonumber \\
& &+\frac{1}{6}\Sigma e^{-2h}(K_1\Gamma_1+K_4\Gamma_4)(\cos\theta\gamma_{01}\Gamma_{12}+\sin\theta\Gamma_2)(\eta\otimes \chi)\nonumber\\
& &-\frac{1}{12}e^{-2f}(K_2\Gamma_2+K_5\Gamma_5)(\cos\theta\gamma_{01}+\sin\theta\Gamma_1)(\eta\otimes \chi)
\end{eqnarray}
which leads to
\begin{eqnarray}
0&=&\frac{1}{2}\left\{h'+\frac{1}{3\sqrt{2}}\Sigma^2(\hat{\xi}^{12}+\hat{\xi}^{45})+\frac{1}{3}\Sigma^{-1}\hat{f}_{345}\right\}\cos\theta (\eta\otimes \chi),\nonumber \\
& &-\frac{1}{2}\left\{\tilde{\kappa}e^{-h}e^{-2i\varphi\gamma_{\hat{r}}}\Gamma_1+\frac{1}{6}\Sigma(2K_1e^{-2h}-K_2e^{-2f})\right\}\sin\theta (\eta\otimes \chi),\nonumber \\
0&=&\frac{1}{2}\left\{-h'+\frac{1}{3\sqrt{2}}\Sigma^2(\hat{\xi}^{12}+\hat{\xi}^{45})-\frac{1}{3}\Sigma^{-1}\hat{f}_{345}\right\}\sin\theta (\eta\otimes \chi),\nonumber \\
& &-\frac{1}{2}\left\{\tilde{\kappa}e^{-h}e^{-2i\varphi\gamma_{\hat{r}}}\Gamma_1+\frac{1}{6}\Sigma(2K_1e^{-2h}+K_2e^{-2f})\right\}\cos\theta (\eta\otimes \chi),\nonumber \\
0&=&(K_5e^{-2f}+2K_4e^{-2h})\sin\theta=(K_5e^{-2f}+2K_4e^{-2h})\cos\theta\, .
\end{eqnarray}
\indent Finally, we consider $\delta\psi_{ri}$ variations which give rise to
\begin{eqnarray}
0&=&\frac{Y'}{Y}(\cos\theta\Gamma_{12}+\sin\theta\gamma_{01}\Gamma_2)\tilde{\epsilon}+\theta'(\cos\theta\gamma_{01}\Gamma_2-\sin\theta\Gamma_{12})
\tilde{\epsilon}\nonumber \\
& &-\varphi'(\cos\theta+\sin\theta\gamma_{01}\Gamma_1)\tilde{\epsilon}+\frac{1}{6}\left\{\frac{1}{\sqrt{2}}\Sigma^2(\hat{\xi}^{12}\Gamma_{12}+\hat{\xi}^{45}\Gamma_{45})-\Sigma^{-1}\hat{f}_{345}\Gamma^{345}\right\}\times \nonumber \\
& &(\cos\theta+\sin\theta\gamma_{01}\Gamma_1)\tilde{\epsilon}-\frac{1}{12}\Sigma e^{-2h}(K_1\Gamma_1+K_4\Gamma_4)(\cos\theta\gamma_{01}\Gamma_{12}+\sin\theta\Gamma_2)\tilde{\epsilon}\nonumber \\
& &-\frac{1}{12}\Sigma e^{-2f}(K_2\Gamma_2+K_5\Gamma_5)(\cos\theta\gamma_{01}+\sin\theta\Gamma_1)\tilde{\epsilon}\, .
\end{eqnarray}
Setting the coefficients of all linearly independent terms to zero, we arrive at the following BPS equations
\begin{eqnarray}
0&=&\left\{\frac{Y'}{Y}+\frac{1}{6\sqrt{2}}\Sigma^2(\hat{\xi}^{12}+\hat{\xi}^{45})+\frac{1}{6}\Sigma^{-1}\hat{f}_{345}\right\}\cos\theta\nonumber \\
& &-\left\{\theta'+\frac{1}{12}\Sigma(e^{-2h}K_1-e^{-2f}K_2)\right\}\sin\theta,\nonumber \\
0&=&\left\{\frac{Y'}{Y}-\frac{1}{6\sqrt{2}}\Sigma^2(\hat{\xi}^{12}+\hat{\xi}^{45})+\frac{1}{6}\Sigma^{-1}\hat{f}_{345}\right\}\sin\theta\nonumber \\
& &+\left\{\theta'-\frac{1}{12}\Sigma(e^{-2h}K_1+e^{-2f}K_2)\right\}\cos\theta,\nonumber \\
0&=&(e^{-2h}K_4+e^{-2f}K_5)\sin\theta=(e^{-2h}K_4+e^{-2f}K_5)\cos\theta,\nonumber \\
0&=&\varphi'\cos\theta=\varphi'\sin\theta\, .
\end{eqnarray}

\end{document}